\documentclass[titlepage,runningheads]{llncs}
\usepackage{hyperref}
\hypersetup{%
    hidelinks,
    pdftitle={SusTrainable: Promoting Sustainability as a Fundamental Driver in Software Development Training and Education},
    pdfauthor={Pieter Koopman, Mart Lubbers and João Paulo Fernandes (Eds.)},
    pdfsubject={Teacher Training Proceedings},
}
    
\usepackage{url}
\usepackage{tikz}
\usetikzlibrary{arrows, shapes, shapes.misc, chains, patterns}

\usepackage{amsmath}
\usepackage{pifont}
\usepackage{subfig}
\usepackage{float}

\usepackage{booktabs}
\usepackage{xcolor}

\usepackage[capitalise]{cleveref} 
    \crefname{sublisting}{listing}{listings}
    \Crefname{sublisting}{Listing}{Listings}

\usepackage[frozencache=true,cachedir=mcache]{minted}
        \definecolor{pygKT}{RGB}{0, 0, 128}
        \definecolor{pygS}{RGB}{0, 0, 255}
        \definecolor{pygW}{RGB}{0, 135.15, 0}
    \setminted{autogobble}
    \setminted{escapeinside=||}
    \setminted{fontsize=\footnotesize}
    \setminted{mathescape}

\newsavebox{\mintedbox}

\def\CPPTitle{\texorpdfstring{C\raise.22ex\hbox{{\footnotesize +}}\raise.22ex\hbox{\footnotesize +}}{C++}}
\newcommand{\CC}{\texorpdfstring{C\nolinebreak\hspace{-.05em}\raisebox{.4ex}{\tiny\bf +}\nolinebreak\hspace{-.10em}\raisebox{.4ex}{\tiny\bf +}}{C++}}

\graphicspath{{picture/}}
\usepackage{tabularx}
\usepackage{color}
\usepackage{listings}

\usepackage{amsfonts, amssymb, amsmath}

\renewcommand{\orcidID}[1]{\href{https://orcid.org/#1}{\textsuperscript{\includegraphics[scale=.5]{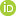}}}}

\title{
SusTrainable: Promoting Sustainability as a Fundamental Driver in Software Development Training and Education\\[2ex]
Teacher Training\\
November 1--5, Nijmegen, The Netherlands\\
Revised lecture notes
}
\author{Pieter Koopman, Mart Lubbers and João Paulo Fernandes (Eds.)}
\date{April 2022}

\begin{document}
\begin{titlepage}
  \let\footnotesize\small
  \let\footnoterule\relax
  \let \footnote \thanks
  \null\vfil
  \vskip 60pt
  \begin{center}%
    {\LARGE
        SusTrainable: Promoting Sustainability as a Fundamental Driver in Software Development Training and Education\\[2ex]
        \large
        Teacher Training\\
        November 1--5, Nijmegen, The Netherlands\\
        Revised lecture notes
    \par}%
    \vskip 3em%
    {\large
     \lineskip .75em%
      \begin{tabular}[t]{c}%
        Pieter Koopman\and Mart Lubbers\and Jo\~ao Paulo Fernandes (Eds.)
      \end{tabular}\par}%
      \vskip 1.5em%
    {\large April 2022 \par}
  \end{center}\par
  \vfil\null
\end{titlepage}
\frontmatter

\chapter*{Preface}
\markboth{Preface}{Preface}

These are the proceedings of the first teacher training of the Erasmus+ project \emph{No. 2020{-}1{-}PT01{-}KA203{-}078646 --- Sustrainable}.
The full title of this project is \emph{Promoting Sustainability as a Fundamental Driver in Software Development Training and Education} and the interested reader may know more about it at:

\begin{center}
\url{https://sustrainable.github.io/}
\end{center}

There is a worldwide consensus that sustainability is a key driver for the development of modern society and the future of our planet.
Nevertheless, experts estimate that the ICT domain consumes currently 6--9\% of the total energy consumption and fear that this will grow to 10--20\% by 2030.
Apart from the direct energy consumption for executing software also the production of software has severe sustainability issues.
Creating and maintaining software is a labour-intensive activity. 
Currently, we are reaching the boundaries of what can be achieved by that available manpower.

To cope with those sustainability challenges, we need to improve our way of making software.
The objective of the Sustrainable project is to train the software engineers of the near feature to deal with these challenges.
The project is executed by a broad and diverse consortium of researchers and educators from 10 selected universities and 7 countries from across Europe.

The flagship contribution is the organization of two summer schools on sustainable software production.
The first summer school is moved due to the Corona pandemic.
It is now scheduled for July 2022 in Rijeka, Croatia.
This summer school will consist of lectures and practical work for master and PhD students in the area of computing science and closely related fields.
There will be contributions from Plovdiv University, University of Ko\v{s}ice, University of Coimbra, University of Minho, E\"otv\"os Lor\'{a}nd University, University of Rijeka, Radboud University Nijmegen, University of Pula, University of Amsterdam and Babe\textcommabelow{s}-Bolyai University.

To prepare and streamline this summer school, the consortium organized a teacher training in Nijmegen, the Netherlands.
This was an event of five full days on November 1--5, 2022 organized by Ingrid Berenbroek, Mart Lubbers and Pieter Koopman on the campus of the Radboud University Nijmegen, The Netherlands.
Sustainability is one of the themes within the strategy of the Radboud University called `A Significant Impact', see \url{https://www.ru.nl/sustainability/organisation/radboud-sustainable/}. Sustainability plays an important role in education, research, and management.

In this teacher training, we discussed and tuned the contents of the planned lectures.
The topics discussed in each contribution are:
\begin{itemize}
\item 
    the contents of the intended lecture, but not the entire lecture itself;
\item 
    what is the intended audience;
\item 
    what is the required knowledge and skills of the audience;
\item 
    what are required materials (hardware and software) for the exercises;
\item 
    what will students know after this lecture and can this be used in subsequent lectures;
\item 
    what is the relation with other contributions, other dependencies and what are the lessons learned during the teacher training.
\end{itemize}
This proceedings demonstrates that the summer school will cover a broad range of topics within sustainability.
Ranging for instance from including sustainability in the planning of software products, via selecting programming languages based on the energy consumption of the generated code, to concrete energy reductions and improving maintainability.
Apart from the contributions listed in this proceedings, there were also a number of presentations related to computing science education and sustainability that will be part of the coming summer school:
\begin{itemize}
\item
    Dr.\ Peter Achten from the Radboud University: Success for a bachelor in computing science. The computing science bachelor is the highest ranked bachelor in the Netherlands. This talk reveals some of the secrets behind this success.
\item
    Dr.\ Bernard van Gastel from the Radboud University Nijmegen: Static energy analysis of software controlling external devices. He showed how one can reason about the energy use of external hardware controlled by software.
\item
    Dr.\ Pieter Koopman from the Radboud University: PersonalProf. This is a tool that can provide almost instantaneous feedback to students handing in programming assignments.
\item
    Dr.\ Jo\~ao Saraiva from the university of Minho From Technical Debt to Energy Debt. Treating an overly high energy use as a software debt enables us to handle such a lack of sustainability as part of software maintenance.
\item
    Dr.\ Cla\'udio Gomes from the university of Porto: On the use of Quantum Computers for Sustainability.	
\end{itemize}
Those presentations provided useful background knowledge for the teaching team.
See \url{https://sustainable.cs.ru.nl/TeacherTrainingNijmegen} for the complete schedule.

The contributions were reviewed and give a good overview of what can be expected from the summer school.
Based on these papers and the very positive and constructive atmosphere, we expect a high quality and pleasant summer school.
We are looking forward to contributing to a sustainable future.

\medskip

\begin{flushright}
\noindent April 2022\hfill
{Pieter Koopman\footnote{\label{ftn:ru}Radboud University Nijmegen, The Netherlands}}\\
{Mart Lubbers\footnotemark[\getrefnumber{ftn:ru}]}\\
{Jo\~ao Paulo Fernandes\footnote{University of Porto, Portugal}}\\
\end{flushright}

\tableofcontents

\mainmatter%

\title{Using Virtual Environments to Provide Context in Verified Software Development}
\author{\v{S}tefan Kore\v{c}ko\orcidID{0000-0003-3647-6855}}
\authorrunning{\v{S}. Kore\v{c}ko}
\institute{Department of Computers and Informatics, 
	Faculty of Electrical Engineering and Informatics,
	Technical University of Ko\v{s}ice, Slovakia\\
	\email{stefan.korecko@tuke.sk}
}
\maketitle
\begin{abstract}
Formal methods for software development allow to verify the correctness of a system with respect to its formally specified properties but cannot verify the properties against informal requirements.  Instead, validation via animation of formal specifications is used. To improve the validation, an animatable formal specification or an executable prototype derived from it can be embedded into a 3D virtual environment, resembling the one where the system will be used. Utilization of the virtual environments instead of real ones contributes to the environmental and economic sustainability by saving resources.  This paper outlines a short course that demonstrates both the verified software development of the executable prototype using a formal method called B-Method and the embedding of the prototype into a web-based collaborative virtual environment for validation purposes.

\end{abstract}
\section{Introduction}
Several formal methods, such as B-Method~\cite{BKnihaAbrial,schneiderBbook}, Event-B~\cite{EventBBook} or VDM~\cite{VDMPPbook}, allow to verify the correctness of a formally specified system with respect to its properties. However, all the effort put into the verification process would be wasted if the properties themselves are incorrect. To validate the properties against informal requirements of the system, animation (execution) of formal specifications is used. The animation is supported by tools, called animators. The animations are usually performed via a command-line interface while some animators also allow simple visualizations.  

Especially in the cases when the software to be developed is a control system of an autonomous entity that performs its duties in a physical space, it is beneficial if the animation, and corresponding validation, occur in a 3D virtual environment that resembles the entity and the environment where it is about to operate. Provided that the virtual representation of the entity is connected to the formally developed software, the properties of the software can be validated in interaction with the virtual environment. The entity may also interact with real users (people), represented by avatars in the virtual environment. Thanks to the recent developments in virtual reality (VR) hardware, VR headsets in particular, a full immersion of a user to the virtual environment can be provided together with a realistic capture of his or her movements. 

In this paper, we present a short course that demonstrates a practical usability of the approach to the verified software validation, described above. The course revolves around a development of a control software for a cleaning robot and a simple virtual environment, where the robot will be placed and perform the cleaning operation. The verified development of the control software will be primarily presented during a lecture part of the course while the practical part will be devoted to a more attractive construction of the virtual environment. The verified software development will use B-Method and the virtual environment will be constructed using A-Frame~\cite{wwwAFrame}, a web-based extended reality (XR) software framework. 
A general description of the approach and its relation to the sustainability field is given in Section~\ref{sec:verified-software-development-with-validation-in-virtual-environments}. The course is outlined in Section~\ref{sec:course-proposal} and in Section~\ref{sec:kconclusion} the paper concludes with remarks on a practical execution of the course including the target group, prerequisites and software, hardware and time requirements.

\section{Verified Software Development with Validation in Virtual Environments}\label{sec:verified-software-development-with-validation-in-virtual-environments}

\begin{figure}[!ht]
	\centering
	\includegraphics[width=0.8\textwidth]{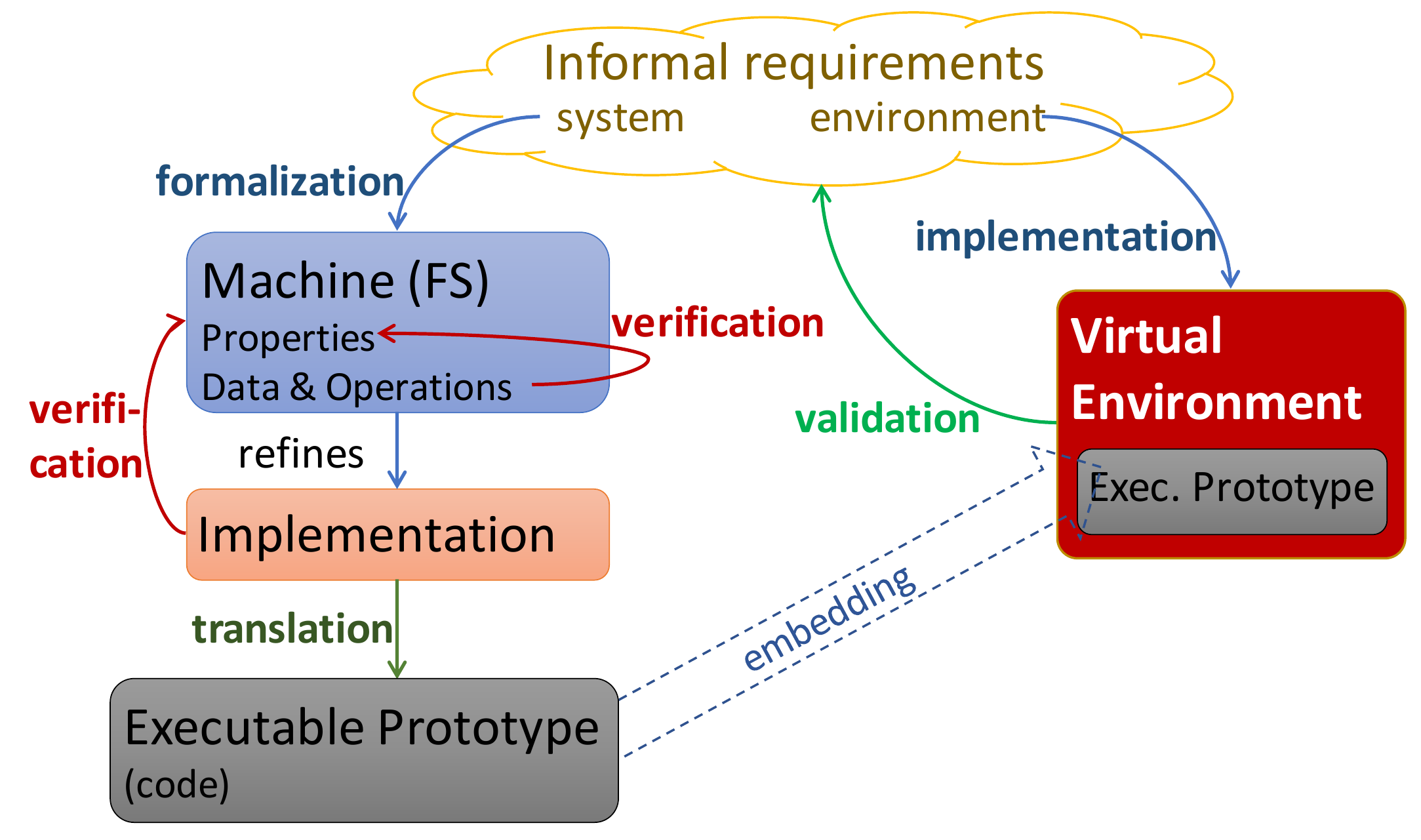}
	\caption{Development process in general}%
	\label{figDevProcess}
\end{figure}

How the verified software development utilizing formal methods (FM) can be assisted by validation in virtual environments is illustrated in Fig.~\ref{figDevProcess}. Fig.~\ref{figDevProcess} assumes the usage of the B-Method as FM, but it can be used with any FM that incorporates formal refinement and generation of source code from implementable specifications.   As it is usual in software development, everything starts with informal requirements on the system to be developed. These informal requirements are specified in a natural language and, therefore, are prone to ambiguous interpretation.  

The first step of the verified software development with B-Method is a creation of a formal specification of the system. The specification consists of components, called machines. A typical machine contains data elements, such as state variables and constants, and operations over the state variables. Each machine with state variables also includes a formula defining invariant properties that should hold in any state of the machine. These formulas, called invariants, should contain a formalization of critical parts of the informal requirements. Whether the invariants really hold in each state is checked during the verification, which is carried out by theorem proving or model checking. The formal specification is then refined in one or more steps into the final, implementable, form consisting of components, called implementations. Whether the refinement preserves the properties defined on the abstract specification level is, again, checked by the verification. A code generator is then used to translate the implementation into a source code in a general-purpose programming language, creating an executable prototype of the developed software.  

In parallel with the software development, the informal requirements are also used to create a virtual environment (VE), where the executable prototype will be validated and evaluated. Tools such as 3D editors or game engines, together with 3D model repositories, should be used to minimize the effort needed for the VE creation. The executable prototype is then embedded into the virtual environment, where it can be validated and evaluated. The purpose of the validation is to check whether  the informal requirements have been properly understood and formalized, that is whether the prototype has the right properties. An additional evaluation or testing may be used to fine-tune parameters of the prototype or testing it under conditions that may not been considered originally. 

According to the established definitions~\cite{goodland1994environmental,goodland1996environmental}, this approach contributes to the environmental and economic sustainability.  This is because it promotes creating virtual representations and prototypes instead of real ones, thus saving valuable resources. And because the validation and evaluation take place in a virtual space, where the participants can connect remotely, it also saves potential traveling costs. Such position of VR with respect to the sustainability is also recognized in~\cite{hamid2014virtual}, where the VR role in manufacturing is seen in product design evaluation and virtual prototyping. There are several implementations of similar approaches to the VR utilization, for example a welding process design~\cite{chen2021Implications} or a disassembly line development~\cite{Rocca2020Integrating}. 

\section{Course Proposal}\label{sec:course-proposal}

The course will demonstrate the outlined process on a case of a verified development, validation and evaluation of a control software of a fictional autonomous cleaning robot. This case has been introduced in~\cite{hudak2020LIRKIS}, subsequently implemented and used in~\cite{korecko2021Experimental} for usability evaluation of related web-based XR components. 

\begin{figure}[!ht]
	\centering
	\includegraphics[width=0.7\textwidth]{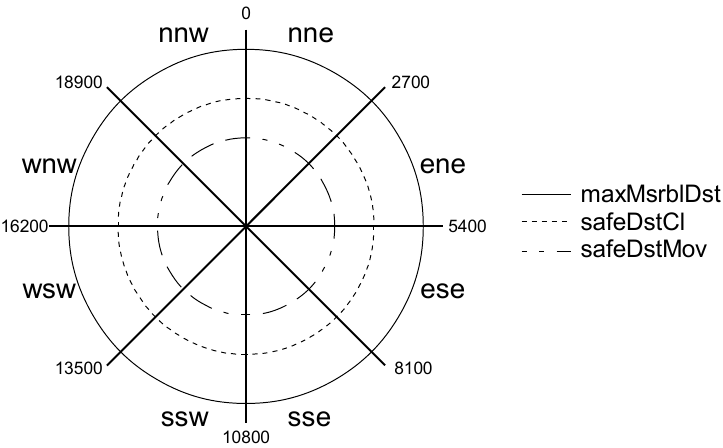}
	\caption{Proximity sensors arrangement of the cleaning robot}%
	\label{fig_ProxSensorsArrangementMin}
\end{figure}

The cleaning robot, called \emph{cBot}, should operate as follows: The robot has a list of locations to clean and it proceeds to clean them one by one. After cleaning all locations, it goes to a parking position and switches to a standby mode. It is assumed that the robot performs its job in areas where people and other living beings occur, so it is critical to prevent it from hurting them. To detect them, the robot has a circular sensor array (Fig.~\ref{fig_ProxSensorsArrangementMin}) with eight sensors, each covering a $45^{\circ}$ ($2700$ minutes) region, north-north-east ({\sffamily nne}) to north-north-west ({\sffamily nnw}).  Each sensor returns a value equal to the distance to a nearest living being, detected in the corresponding region. If no living being is detected in the region, then the returned value is equal to the maximum distance, measurable by the array ({\sffamily maxMsrblDst} in Fig.~\ref{fig_ProxSensorsArrangementMin}). The robot may hurt someone when moving or cleaning. According to this, we can define two important values 

\begin{itemize} 
	\item 
	{\sffamily safeDstCl}, the minimum safe distance from the robot when cleaning and 
	\item 
	{\sffamily safeDstMov}, the minimum safe distance from the front of the robot when moving
\end{itemize} 
and introduce two safety critical properties the robot has to follow: 

\begin{description} 
	
	\item[SP.1]  
	
	The cleaning cannot start or continue if anyone (i.e.\ any living being) gets as close or closer to the robot as {\sffamily safeDstCl}. 
	
	\item[SP.2]  
	
	The robot cannot move if anyone gets as close or closer to its front as  {\sffamily safeDstMov}. 
	
\end{description} 
As it can be seen in Fig.~\ref{fig_ProxSensorsArrangementMin}, we also define that {\sffamily safeDstMov} $\leq$ {\sffamily safeDstCl} $\leq$ {\sffamily maxMsrblDst}. And for the sake of simplicity, we assume that the robot operates in a perfectly flat environment and all living beings move on the ground and are high enough to be detected by the array. This means that their vertical position does not need to be considered. 

\begin{figure}[!ht]
	\centering
	\includegraphics[width=0.9\textwidth]{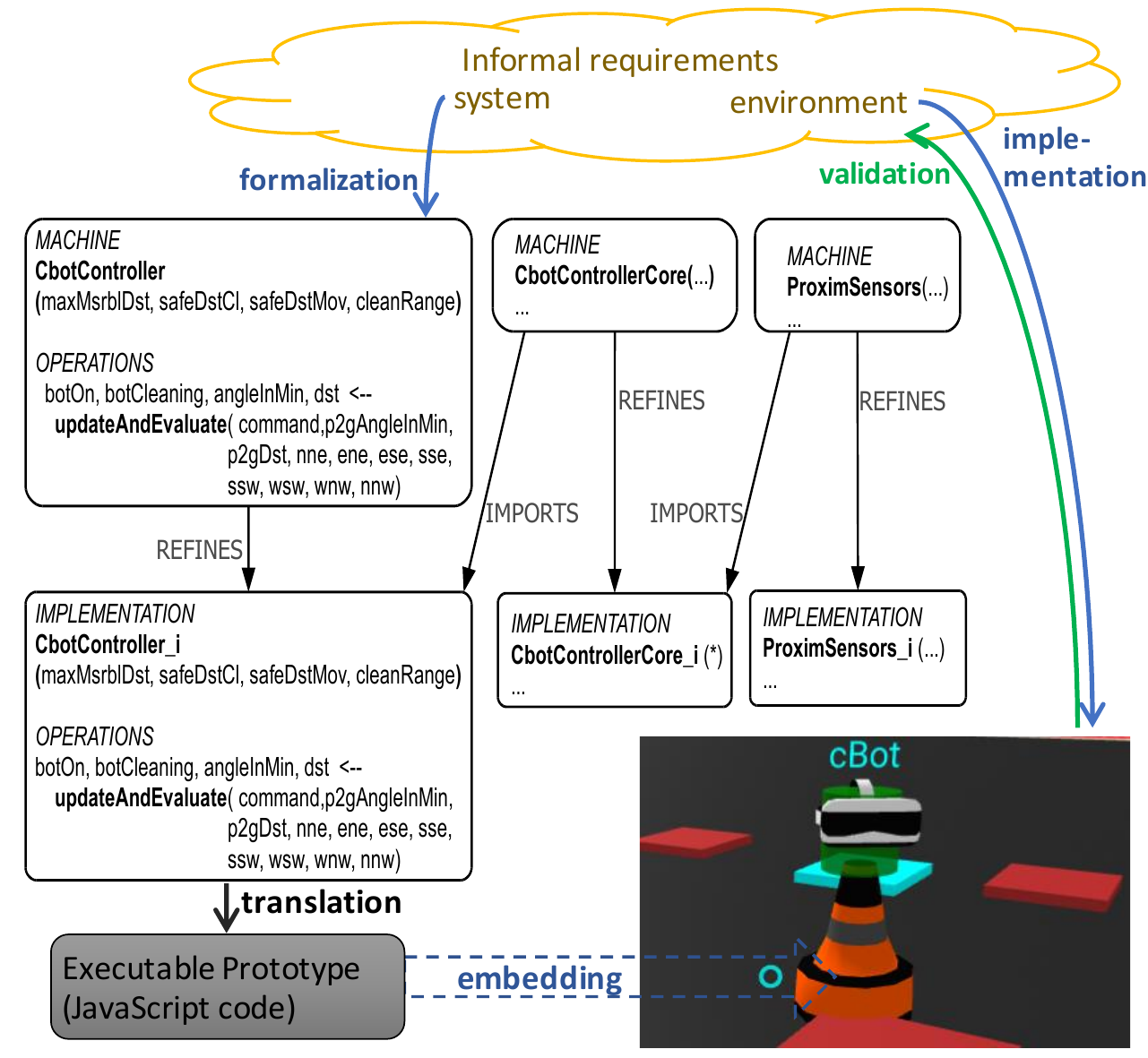}
	\caption{Development process as used in the course}%
	\label{figDevProcessCBot}
\end{figure}

How the course will utilize the process from Fig.~\ref{figDevProcess} for the \emph{cBot} case is shown in Fig.~\ref{figDevProcessCBot}. The purpose of the verified control software, developed using B-Method, will be to command the robot according to the desired action and readings from the sensor array. The formal specification of the software will consist of three machines, {\sffamily CBotController}, {\sffamily CBotControllerCore} and {\sffamily ProximSensors}. These will be refined in one step to the corresponding implementations ({\sffamily CBotController\_i}, etc.). The machine {\sffamily ProximSensors} (and its implementation) contain operations for reading and interpreting the sensor values. The operations of {\sffamily CBotControllerCore} compute the commands for the robot and the only operation of {\sffamily CBotController}, the {\sffamily updateAndEvaluate}, serves as the interface of the whole controller. The {\sffamily updateAndEvaluate} operation takes a desired action for the robot (the input parameter {\sffamily command}), a compass angle ({\sffamily p2gAngleInMin}) and distance ({\sffamily p2gDst}) to the position where the robot should go and the sensor array readings ({\sffamily nne} to {\sffamily nnw}). Then it evaluates the desired action and destination with respect to the sensor readings and returns instructions, which the robot will follow. These include a decision to turn on or switch the robot to standby (the output parameter {\sffamily botOn}), to start or stop the cleaning process ({\sffamily botCleaning}) and a compass angle ({\sffamily angleInMin}) and distance ({\sffamily dst}) to a position where the robot will go. The command may be to switch to standby mode immediately (value {\sffamily 0}) or go to the desired destination and then clean ({\sffamily 1}), do not clean but stay on ({\sffamily 2}) or switch to standby ({\sffamily 3}). The verified development will take place in the Atelier-B~\cite{wwwAtelierB} integrated development environment and prover. After the development of the controller is completed, it will be translated to JavaScript by a modified version of the BKPI compiler~\cite{korecko2011BKPI} and embedded to a virtual environment. 

The virtual environment will be web-based, which means that it runs directly inside a web browser. It will be implemented using A-Frame~\cite{wwwAFrame}, a software framework for XR web applications. The environment will contain a stylized representation of the \emph{cBot}  and avatars of other users. The places to clean will be represented by red tiles. The robot and the tiles can be seen in the right bottom corner of Fig.~\ref{figDevProcessCBot}. An extension of A-Frame, called Networked-Aframe~\cite{wwwNAF}, will be used to allow multiple users to share the same environment. The environment will also contain a JavaScript code, which will operate the robot and call the {\sffamily updateAndEvaluate} operation of the embedded controller periodically. The code will also translate positions of the avatars of other users to the readings of the sensor array. To obtain the exact positions of the avatars, it will utilize another extension of A-Frame, called Enhanced Client Access layer~\cite{hudak2020LIRKIS,korecko2021Experimental}.

\section{Conclusion}\label{sec:kconclusion}
In this paper, we outlined a course that demonstrates how the verified software development can be connected with validation and evaluation in virtual environments. The course is intended for all audiences with a basic knowledge of computer programming and web technologies, namely HTML and JavaScript. 

The course covers two distinct themes; verified software development with formal methods and creation of virtual environments with web technologies.  Provided that the audience is primarily interested in only one of these themes, the course can be carried out within 4 to 8 hours. In such case, the less interesting theme will be covered by a lecture while a hands-on experience will be provided for the more interesting one. A form of the course with practical experience in both themes requires about 20 hours. Ordinary personal computers with Internet connectivity and running Windows operating system, Atelier-B and a web browser are all what is needed for the course. While not necessary, some VR headsets, such as Oculus Quest, should be available during the course. 

After the course, it is expected that the participants will understand the role of formal methods in software development and the difference between validation and verification. They will also know how to create web-based virtual environments and how the utilization of such environments can contribute to the sustainability. 

\section*{Acknowledgement}
This paper acknowledges the support of the Erasmus+ Key Action 2 (Strategic partnership for higher education) project No. 2020-1-PT01-KA203-078646: ``SusTrainable - Promoting Sustainability as a Fundamental Driver in Software Development Training and Education''.
The information and views set out in this paper are those of the author(s) and do not necessarily reflect the official opinion of the European Union. Neither the European Union institutions and bodies nor any person acting on their behalf may be held responsible for the use which may be made of the information contained therein.

\title{An Overview of Multicriteria Decision Methods}
\author{Lu\'{\i}s Paquete\orcidID{0000-0001-7525-8901}}
\institute{CISUC, Department of Informatics Engineering, \\ University of Coimbra, Portugal\\
\email{paquete@dei.uc.pt}}
\maketitle
\begin{abstract}
Multicriteria decision methods aim at supporting decisions when several
conflicting points of view need to be considered.  This is often the case in
the evaluation of alternative software engineering projects, for which not only
cost/profit matters, but also other less quantifiable criteria, such as team
well-being and project \emph{sustainability}. This class will provide an
overview of the main multicriteria decision methods that can be found in 
the literature and discuss their weaknesses and strengths. 
\end{abstract}
\section{Motivation}

Decision problems arise very often in our daily life. For instance, when choosing
a path from work to home, one may take into account not only distance but also
sightseeing quality, road traffic, or some enviromental aspect.  One has to
deal with several possible choices that are, to some extent, incomparable
because of the multiple-criteria nature of the underlying problem.

Software industry also deals with several criteria at different stages
of the software engineering process. For example, the choice for the most
appropriate software engineering practice must take into account whether, or
not, it allows for agile development, customer interaction, early
deployment, risk reduction, continuous testing, sustainability policy
pursuance, and many other criteria.  

Multicriteria Decision Aiding (MCDA), a field of knowledge that arose more
50 years ago, provides formal methods that allow
to take informed decisions. However, in despite of its wide use in many
industrial applications, it is not yet a current practice at software industry
and it is not even taught at graduate IT courses. Application of these methods 
can only be found in specialized literature, 
see~\cite{AbdulwarethA21,BARCUS2008464,BuyukozkanR08,LAI2002134,OteroOWQ10,SanthanamK95,TrendowiczK14,wangL03a}.
For instance, Pereira et al.~\cite{Pereira21} 
rank programming languages with respect to several criteria, such
as energy and memory. Also, some of the problems that arise in Search-Based
Software Engineering are shown to be naturally multicriteria; see survey of
Yao~\cite{Yao13}. Given the increasing relevance of sustainability as
a criterion in the several steps of both software development and 
maintenance processes, it is more
important than ever that software engineers know how to apply multicriteria
decision methods.

\section{Outline}

The main goal of the  class ``An Overview on Multicriteria Decision
Methods'' is to teach students on how to take complex decisions under multiple 
criteria. The underlying scenario in the class is that a Decision Maker has to
take a choice from a finite set of alternatives, each of which is evaluated on
the basis of a finite family of evaluation criteria. These criteria represent
different aspects of evaluation of the alternatives.
It is assumed that each criterion is associated with a monotone direction of
preference. For example, the cost of a software project has a decreasing
direction of preference, but sustainability has an increasing direction of
preference. Some aspects that will be covered in the class are described next.

Typically, decision aiding problems can be recast as choice, ranking or sorting problems. 
In \emph{choice problems}, the goal is to select the best alternative or the best 
subset of alternatives by rejecting those that are worse~\cite{malekmohammadi11}.
In \emph{ranking problems}, all alternatives are ranked from the best to the the worst,
allowing for ties and incomparabilities. In \emph{sorting
problems}~\cite{Morais14}, each alternative is assigned to one or more classes
that have been pre-defined and ordered from the best to the worst according to
some Decision Maker's model.  

In order to deal with any of the three problems above, a performance table is 
constructed, which gives objective information about each alternative 
according to the several criteria at hand.
From this information it is possible to extract a set of interesting
alternatives based on the concept of \emph{dominance relation}; an alternative
$a$ dominates $b$ if and only if $a$ is at least as good as $b$ for all
criteria and better for at least one of them. This is the most basic assertion
of preference with respect to the set of alternatives considered in 
decision problem. 
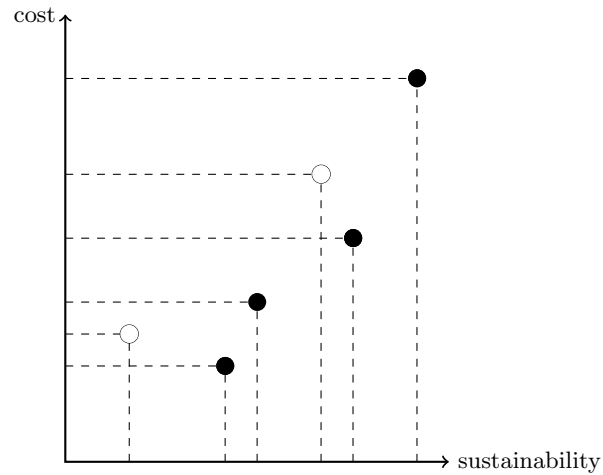
\begin{figure}[t]
	\begin{center}
	\begin{tikzpicture}[scale=1.7]
        \draw [<->,thick] (0,3.5) node (yaxis) [left] {\small cost }
                        |- (3,0) node (xaxis) [right] {\small sustainability};

        \draw[dashed] (5.0/4, 0.0) coordinate (a1) -- 
                      (5.0/4, 3.0/4) coordinate (a2);
        \draw[dashed] (0.0   , 3.0/4) coordinate (a3) -- 
                      (5.0/4, 3.0/4) coordinate (a4);
        \coordinate (aaaa) at (intersection of a1--a2 and a3--a4);
        \fill[black] (aaaa) circle (2pt);
        \draw[dashed] ( 9.0/4, 0.0) coordinate (a1) -- 
                      ( 9.0/4, 7.0/4) coordinate (a2);
        \draw[dashed] (0.0   , 7.0/4) coordinate (a3) -- 
                      ( 9.0/4, 7.0/4) coordinate (a4);
        \coordinate (a11) at (intersection of a1--a2 and a3--a4);
        \fill[black] (a11) circle (2pt);
        \coordinate (a13) at (intersection of a1--a2 and a3--a4);
        \fill[black] (a13) circle (2pt);
        \draw[dashed] ( 11.0/4, 0.0) coordinate (a1) -- 
                      ( 11.0/4, 12.0/4) coordinate (a2);
        \draw[dashed] (0.0   , 12.0/4) coordinate (a3) -- 
                      ( 11.0/4, 12.0/4) coordinate (a4);
        \coordinate (azzz) at (intersection of a1--a2 and a3--a4);
        \coordinate (a) at (intersection of a1--a2 and a3--a4);
        \fill[black] (a) circle (2pt);
        \draw[dashed] ( 2.0/4, 0.0) coordinate (a1) -- 
                      ( 2.0/4, 4.0/4) coordinate (a2);
        \draw[dashed] (0.0   , 4.0/4) coordinate (a3) -- 
                      ( 2.0/4, 4.0/4) coordinate (a4);
        \coordinate (a) at (intersection of a1--a2 and a3--a4);
        \draw[black] (a) circle (2pt);
        \fill[white] (a) circle (2pt);
        \fill[black] (aaaa) circle (2pt);
        \fill[black] (a11) circle (2pt);
        \fill[black] (azzz) circle (2pt);
        \draw[dashed] ( 6.0/4, 0.0) coordinate (a1) -- 
                      ( 6.0/4, 5.0/4) coordinate (a2);
        \draw[dashed] (0.0   , 5.0/4) coordinate (a3) -- 
                      ( 6.0/4, 5.0/4) coordinate (a4);
        \coordinate (a) at (intersection of a1--a2 and a3--a4);

        \fill[black] (a) circle (2pt);
	\draw[dashed] ( 8.0/4, 0.0) coordinate (a1) -- 
                      ( 8.0/4, 9.0/4) coordinate (a2);
        \draw[dashed] (0.0   , 9.0/4) coordinate (a3) -- 
                      ( 8.0/4, 9.0/4) coordinate (a4);
        \coordinate (a) at (intersection of a1--a2 and a3--a4);
        \draw[black] (a) circle (2pt);
	\fill[white] (a) circle (2pt);
\end{tikzpicture}
	\end{center}
\caption{Evaluation of software projects in terms of cost and sustainability}%
\label{mcda:f:1}
\end{figure}

Note that there might not exist a single \emph{dominating} alternative, but a
set of them that are noncomparable.
Figure~\ref{mcda:f:1} illustrates a hypothetical example of six software projects
that are evaluated in terms of two criteria: cost and sustainability. The black
points correspond to projects that are dominating whereas the white correspond
to dominated projects.  A Decision Maker would never choose any project corresponding to a
white point as each of which is dominated by at least one other project
corresponding to a black point. For instance, the most costly of the dominated
projects is dominated by the second most sustainable project.

Although assertive, dominance relation has little discriminant power. 
There might exist too many dominating projects for the Decision Maker to choose from. 
In order to reduce the number of alternatives considerably, there is the need to 
aggregate the several criteria according to some \emph{preference model}.
Three aggregation methods can been considered in the MCDA literature:
\begin{itemize}
	\item \emph{Value function}, which consists of
		assigning to each alternative a real number that is
		representative of its assessment in the
		problem~\cite{Keeney76}. This is the case of the weighted-sum model and 
		goal programming. 
	\item \emph{Outranking relation}, which is a binary relation that takes into
		account the number of criteria in favor, and not in favor,
		of an alternative over another~\cite{Roy96}. This is the basis of
		PROMETHEE~\cite{Brans85} and ELECTRE methods~\cite{Figueira13}.

	\item \emph{Decision rules}, which are logical statements that define 
		some threshold requirements on the chosen criteria and provide
		a recommendation for an alternative that satisfies those requirements~\cite{Greco01}.
\end{itemize}

A further aspect to be mentioned in the class is the concept of representation
in MCDA:\ a representation of a set of noncomparable alternatives is a subset
that optimizes a given \emph{representation quality}~\cite{Sayin00}. This quality
relates to some property of interest that this subset should have in the
criteria space, such as, (i) \emph{uniformity}, the representation points are
as spread as possible; (ii) \emph{coverage}, the representation points are
close enough to the remaining points. The problem of finding the best
representation with respect to each of the properties above can be recast as an
optimization problem, which is NP-hard in general but polynomially solvable
for two criteria~\cite{Vaz15}. 

\section{Learning Goals and Requirements}

Of particular interest for the student is to understand, for a given decision
problem, how multicriteria decision methods can be formalized and how to arrive
to one or more choices. She should be able to relate the several approaches with the
basic notion of dominance, and discuss advantages and disadvantages of each
approach.  

A possible group assignment is to apply the multicriteria decision methods 
discussed in the class to real-world data, such as those provided in
Pereira~\cite{Pereira21}. These data contain time, memory and energy spent
by 10 programs implemented in 27 different programming languages.
In the context of the assignment, each programming language is an alternative
that is evaluated with respect to different criteria such as time, memory and
energy.  The assignment would consist of selecting from these data the best (or
to rank the) programming
languanges according to a given application and the preferences expressed by
the students.

There is no particular skill required for this class. Some knowledge
on binary relations might be useful, but this can also be
provided in the class.  The exercises can be solved with pencil and paper (and
rubber), and in the case of more complex exercises with a simple spreadsheet
application.  Several textbooks and surverys are available that might be useful
for those that would like to deepen their knowledge on Multicriteria Decision
Methods
such as Corrente et al.~\cite{Corrente21}, Roy~\cite{Roy96}, Ehrgott et
al.~\cite{Ehrgott10} and Keeney~\cite{Keeney76}.

\subsubsection{Acknowledgements} 
This work is financed by national funds through the FCT – Foundation for
Science and Technology, I.P.  within the scope of the project CISUC –
UID/CEC/00326/2020. 
This paper acknowledges the support of the Erasmus+ Key Action 2 (Strategic
partnership for higher education) project No. 2020{-}1{-}PT01{-}KA203{-}078646:
SusTrainable --- Promoting Sustainability as a Fundamental Driver in Software
Development Training and Education.
L.~Paquete wishes to thank J.R.~Figueira from the University of 
Lisbon for providing part of the material to be used in this class.

\title{Save the Earth, Program in C++!}
\author{Zolt\'{a}n Porkol\'{a}b\orcidID{0000-0001-6819-0224} \and
Rich{\'{a}}rd Szalay\orcidID{0000-0001-5684-5158}
}
\authorrunning{Z. Porkol\'{a}b \and R. Szalay}
\institute{Department of Programming Languages and Compilers,\\
Institute of Computer Science, Faculty of Informatics,\\
Eötvös Lor\'{a}d University,\\ Budapest, Hungary\\
\email{\{gsd,szalayrichard\}@inf.elte.hu}}
\maketitle
\begin{abstract}
Green computing --- paying attention to energy consumption of programs --- 
is getting more and more important in software construction. When dealing with 
high performance systems, cost factors of energy consumption and cooling are 
important. When programming small, or embedded devices the battery
capacity is the restriction. Earlier research show that the C and C++ 
programming languages are ideal choices when we decide based on the 
performance/energy consumption ratio. However, to exploit these 
possibilities the programmer must deeply understand the corresponding 
language constructs and should apply them in the most appropriate way.
In this paper we describe our efforts to teach energy efficiency for 
Master students at E\"{o}tv\"{o}s Lor\'{a}nd University. As part of the 
\emph{Advanced C++ programming} course we teach language techniques like 
the move semantics, RAII (resource acquisition is initialization) technique, 
effective concurrent programming as well as some compiler optimization methods. 
Understanding how students can write programs while maximizing performance, 
students will be able to create software with minimal energy consumption.

\end{abstract}

\section{Introduction}%
\label{intro}

Energy efficiency is an important part of Informatics which comes more and 
more into the focus of interest. In the recent years, the growing use of mobile 
phones and \emph{Internet of Things} (IoT) devices have shown that the decrease
of energy consumption can improve the user experience (with longer time between
charging) and can extend the lifetime of tools, where the re-charging or 
change of battery is not a feasible option~\cite{pinto:newconcern}. The life 
expectancy of a Mars rover is highly dependent on its energy consumption
whether we are speaking of non-replaceable radioactive sources or degrading
and less and less effective solar panels. On the other end of applications 
--- high frequency trading or large mathematical simulations --- energy 
consumption causes problems due to the generated heat which requires 
complex and expensive cooling.

Lower energy consumption does not depend only on hardware but also on the 
software. The programming languages, the data structures, the algorithms 
used by the programmer have serious effect on the energy efficiency.
Green software, therefore, is recently a very attractive research area, 
where various ``energy smells'' and good practices are 
discussed~\cite{pereira:collections,pereira:influence,linares:greedy}. 
At the same time, surveys revealed that the programmers have a limited 
knowledge of these practices~\cite{pang:know}. Therefore, there is an 
increased demand to introduce energy awareness in software technology
education~\cite{cai:undergraduate,sammalisto:sustain,turkin:masters,xiong:higher,saraiva:cscurriculum}.

Earlier studies have shown, that the C and C++ programming languages perform
extremely well concerning energy efficiency~\cite{rua:metrics,pereira:languages}. This is the consequence of various effects. The C and C++ programming
languages were designed for flexibility and efficiency. While during the 
more than 40 years of development the original goals of C++ have been 
refined, efficiency and flexibility have been maintained without 
compromise~\cite{stroustrup:evolving}. C++ does not use a virtual machine, its 
source is compiled directly to machine code of the target architecture which
lets the compiler perform extreme optimizations. The language has features to 
directly exploit hardware possibilities using pointers, the address operator,
and other similar features. Although there exist a few complex and ineffective 
language elements in C++ (like virtual functions, dynamic cast), these are
almost never used in the standard library and most of the situations can be 
avoided by the programmers in performance critical applications.

In this paper, we describe how we teach efficiency concerns and energy 
awareness in the \emph{Advanced C++ programming} course at the E\"{o}tv\"{o}s 
Lor\'{a}nd University. In Section~\ref{prog}, we overview the programming 
language education at the Faculty of Informatics, and the prerequisites of 
the course. In Section~\ref{advanced}, we describe the course and its 
important components regarding energy efficiency. Our paper concludes 
in Section~\ref{conclusion}.

\section{Programming languages courses at E\"{o}tv\"{o}s Lor\'{a}nd University}%
\label{prog}

In accordance with the Bologna framework~\cite{eu:bologna}, in Hungary the 
higher level education on informatics is separated into the Bachelor (B.Sc.), 
the Master (M.Sc.) and the Doctorate (Ph.D.) levels. Each level is designed 
to have a coherent conglomeration of lectures providing a solid knowledge and 
skill set at the exit. Courses have mandatory and suggested prerequisites 
based both on the theoretical and practical knowledge required to successfully
accomplish the topic.

Within the Faculty of informatics of E\"{o}tv\"{o}s Lor\'{a}nd University,
the Department of Programming Languages and Compilers is responsible for 
teaching the most fundamental programming languages. Languages to teach 
were selected both for the purpose of theoretical and practical reasons. 
They should provide the fundamentals of the most prevalent programming
paradigms, like procedural, object-oriented and functional ones. They should
demonstrate the most important programming language elements, like scope and
lifetime rules, defining and calling subprograms, the various parameter passing
methods, error handling, etc. The same time they should serve as tools for
non-language related subjects: networking, database knowledge, etc.
Finally, ideally they should fit to the time restrictions of the semester: 
13 lectures in 90{-}120 minutes and the same number of practical sessions 
should be enough the teach the fundamentals, together with some individual
activities of the students. 

The programming languages we teach at the very first semester are the Haskell 
programming language~\cite{peytonjones:haskell} as a functional language and 
the C programming language~\cite{ansi:c} as a fundamental imperative language. 
Choosing the C language was according to the previously described principle: 
it is the most minimal procedural language where you can teach all 
fundamentals, from pointers and low level memory access to scope and lifetime 
rules, to modular design. We consider it essential for the future software 
engineers to understand the basics of hardware architecture and its 
use~\cite{drepper:memory}. Students may have an easier task to write correct 
programs in languages with advanced memory handling, having features like 
garbage collection and managed pointers, but after learning Java or C\# as the 
first language it is almost impossible to teach them the low level requirements
essential to write energy efficient and high performance software.
C also serves as a good base for further curricula: its syntax and command 
structure sets the foundation of Java and C++. Although C is not perfect 
(neither of the languages would be) it serves as a good foundation for 
further courses. Such a follow-up is the obligatory Java programming language
course in the second semester to lay out object-oriented fundamentals.

The \emph{Basic C++} course is an elective course in the third semester 
following the obligatory Java course. During the course, students learn the 
differences between C and C++, including the standard C++ library elements: the 
input/output system, the \texttt{std::string} class~\cite{iso:cpp}. We place 
great emphasis on generic programming~\cite{stepanov:stl} and the Standard 
Template Library (STL): the most important container classes, the standard 
algorithms and the concepts of iterators~\cite{meyers:stl}. 

It is important to mention, however, that this basic C++ course is mostly 
restricted to pre-C++11 language elements. C++11 has introduced a high number 
of new language features, making C++11 almost a completely new programming 
language. Teaching lambda functions, smart pointers, move semantics, variadic 
templates, the use of the \texttt{auto} keyword and similar elements simply 
does not fit the time frame of a single semester. Another consideration is, 
that although most of these new features are highly valuable, making the 
language safer and more expressive, unaware use of these new features may 
lead to unexpected behavior and dangerous bugs~\cite{meyers:modern}.

\section{The Advanced C++ programming course}%
\label{advanced}

The \emph{Advanced C++ programming} course is a 5 ECTS credit subject for 
Master students in Hungarian language. Its main goal is to teach the 
professional methods of the C++ programming language: core language features 
and standard library elements. Although there is no formal prerequisite of 
this course, we strongly suggest the students to complete the elective 
Basic C++ course in BSc.\ we discussed in Section~\ref{prog}. Albeit the 
Advanced C++ programming is a Master course, it is a subject frequently taken 
by Bachelor students who want to gain proficiency in the C++ language. As an 
average only one of out five graduated BSc.\ students continue learning on MSc. 
level; many Bachelor students take this lecture to build up industrially 
relevant knowledge before they leave the University.

The predecessor of the Advanced C++ programming lecture was a 3 ECTS course 
called \emph{Multiparadigm programming}. As its title tells, this lecture used 
advanced C++ techniques to demonstrate the different programming paradigms:  
functional paradigm (lambda functions, template metaprograms), generic
programming (templates, advanced STL), concurrent programming (C++11 memory
model) and how these paradigms cooperate. In Advanced C++ programming, the 
course has been extended by practice sessions and lab works, where students 
can improve their practical programming skills as well as their abilities 
to design complex, high-level C++ libraries.

During the course we put emphasis on writing effective and energy efficient 
code. The following is discussed in details.

{\bf Constexpr}
The evaluation of trivial expressions where all arguments are given in compile 
time is done by the compiler for many programming languages. However, the 
evaluation of more complex expressions, especially function calls are almost 
always done in run time. Since C++11, and even more since C++14, the programmer
can define \texttt{constexpr} variables and functions. Constexpr language
elements may be evaluated in compile time (if all of the necessary components 
is known in compile time) or can be executed run time otherwise. With 
constexpr the programmer can transfer computation to compile time, therefore 
can save energy and execution time in run time. Also, more maintainable code
leads to more sustainable software maintenance~\cite{stroustrup:thriving}
practices.

{\bf Move semantics}
The C++ programming language by default uses \emph{value semantics}, i.e.
variables, function parameters are representing the data directly, without 
any indirection, like a reference or pointer. Other languages, especially 
Java or Python, however, use \emph{reference semantics}: here a variable 
is only represents a reference or pointer which refers to the data.
The difference is the most straightforward when we examine a single assignment
statement: \verb|x = y| which means in C++ to copy the bytes of \texttt{y} to
\texttt{x}, while in reference semantic languages it means that \texttt{x}
points (refers) to the same value as \texttt{y}. 
While value semantics has the advantage being a more explicit ownership 
strategy, and therefore better optimization possibilities, it has a tendency
to create unnecessary temporary objects in certain situations. This may 
lead to a significant waste of memory and time and cause bad energy 
characteristics of the software. Move semantics~\cite{hinnant:move,siek:move},
introduced in C++11, is a technique to solve this problem, and gives 
a method to the programmer to save energy and other resources.

{\bf Effective variadic templates}
Variadic templates appeared in C++11 to implement generic functions with
variadic type and number of parameters. The typical implementation of such 
functions uses recursion, thus generates a chain of functions with decreasing
number of parameters --- similarly as it happens in many functional programming
languages. Although such intermediate functions may be optimized
out during compilation, there are more direct ways to eliminate them from
the program when we apply clever initialization-list techniques or C++17 
fold expressions. This usually leads to more energy efficient code both at 
compile and run time.

{\bf Effective multithreading}
Writing correct race- and deadlock-free multithreaded applications is always
a challenging task for the students and inexperienced junior programmers.
The C++ memory model~\cite{williams:action}, defined by the C++11 standard
brings solid foundations for writing safe multithreaded programs in C++. 
The default memory model, \emph{sequential consistency}, guarantees a
relatively easy-to-understand method to handle concurrency for the cost
of sometimes unnecessary cache operations. However, cache handling is one
of the major factor of energy consumption. Applying more refined memory
models, like the \emph{acquire-release} or the \emph{relaxed} model, using
atomic variables and operations can reduce cashing activity and therefore
can save significant energy during program execution.

Apart from the theoretical lessons, there is a semester-long homework for 
the students to practice the knowledge acquired in theory. A typical task 
is to implement an STL-like template container with the necessary interface. 
We require the students to apply the most modern techniques including the 
ones connected with energy efficiency.

\section{Conclusion}%
\label{conclusion}

In this paper we reported our efforts to incorporate energy efficiency 
criteria into the Advanced C++ Programming course for informatics at the 
Masters level in the Faculty of Informatics of E\"{o}tv\"{o}s Lor\'{a}nd 
University, Budapest. Sustainability is taking an increasingly important 
place in education. The C++ programming language has been proved being an 
efficient tool in which experienced programmers can write highly effective 
programs with low energy consumption. However, this requires the programmer 
to fully understand some of the key techniques provided by the language. 
Through the Advanced C++ Programming course we deliver both theoretical 
and practical examples for writing efficient, energy aware code. During the
semester students have the opportunity to implement a library component
where they can utilize the knowledge they obtained.

\section*{Acknowledgements}
This paper acknowledges the support of the Erasmus+ Key Action 2 (Strategic 
partnership for higher education) project No. 2020{-}1{-}PT01{-}KA203{-}078646: 
``SusTrainable --- Promoting Sustainability as a Fundamental Driver in Software 
Development Training and Education''. The information and views set out in 
this paper are those of the author(s) and do not necessarily reflect the 
official opinion of the European Union. Neither the European Union institutions
and bodies nor any person acting on their behalf may be held responsible 
for the use which may be made of the information contained therein.

\title{Soft Computing for Sustainability Science}
\author{Goran Mau\v{s}a\orcidID{0000-0002-0643-4577}}
\authorrunning{G. Mau\v{s}a}
\institute{University of Rijeka, Faculty of Engineering, Vukovarska 58, 51000 Rijeka, Croatia 
\email{gmausa@riteh.hr}\\
}
\maketitle
\begin{abstract}

Soft Computing methodologies with their inherent tolerance to uncertainty and imprecision play an important role in addressing many of the challenges in modern data mining frameworks. 
With data quality being of utmost importance in such applications, this tutorial is focused on data cleaning techniques that may solve typical errors and dimensionality reduction techniques that may significantly improve the algorithmic efficiency.
The goal is to provide an overview of the most popular techniques, demonstrate their usage and discuss their sustainability potential.

\end{abstract}

\section{Introduction}


There is no agreed definition of what software sustainability means and how it can be measured, demonstrated or achieved~\cite{venters2014software}.
Sustainability science, on the other hand, is the established research field that seeks to understand the fundamental character of interactions between nature and society. It deals with the challenging problems characterized by high degree of complexity, wide range of uncertainty and little available information. With the aim to enable decision-makers and stakeholders to make and evaluate decisions based on this type of data, it is in constant search for new methodological approaches, tools and techniques.
The complexity related challenges that arise from combining many sources of knowledge, as in sustainability science, require the help of soft computing techniques due to their robustness and adaptability~\cite{Corona2018}. 

Empirical studies have shown that adjusting the coding practices can yield significant energy reduction of basic programming patterns like handling loop counters, initialization and object handling, or using collections~\cite{rocheteau2014green,longo2019reducing,pinto2016comprehensive,oliveira2021,lima2016}.
Hence, it is reasonable to assume that algorithms of high complexity, like soft computing, can considerably improve their efficiency with appropriate practices, too.
The principles of green data mining mainly turn to data, as its most valuable part, and to the models that are being built upon that data~\cite{schneider2019principles}.
The \textit{green} design principles are recognized in the stages of: 
(1) data collection and warehousing, (2) data pre-processing, and (3) model training and fine-tuning. 
While the first stage is often out of reach for data scientists, it is their task to prepare the purified data and chose the appropriate training approach by opting for the simplest solutions and avoiding exhaustive search.

\section{Applied Soft Computing}

\begin{figure}
\centering
\includegraphics[width=0.8\textwidth]{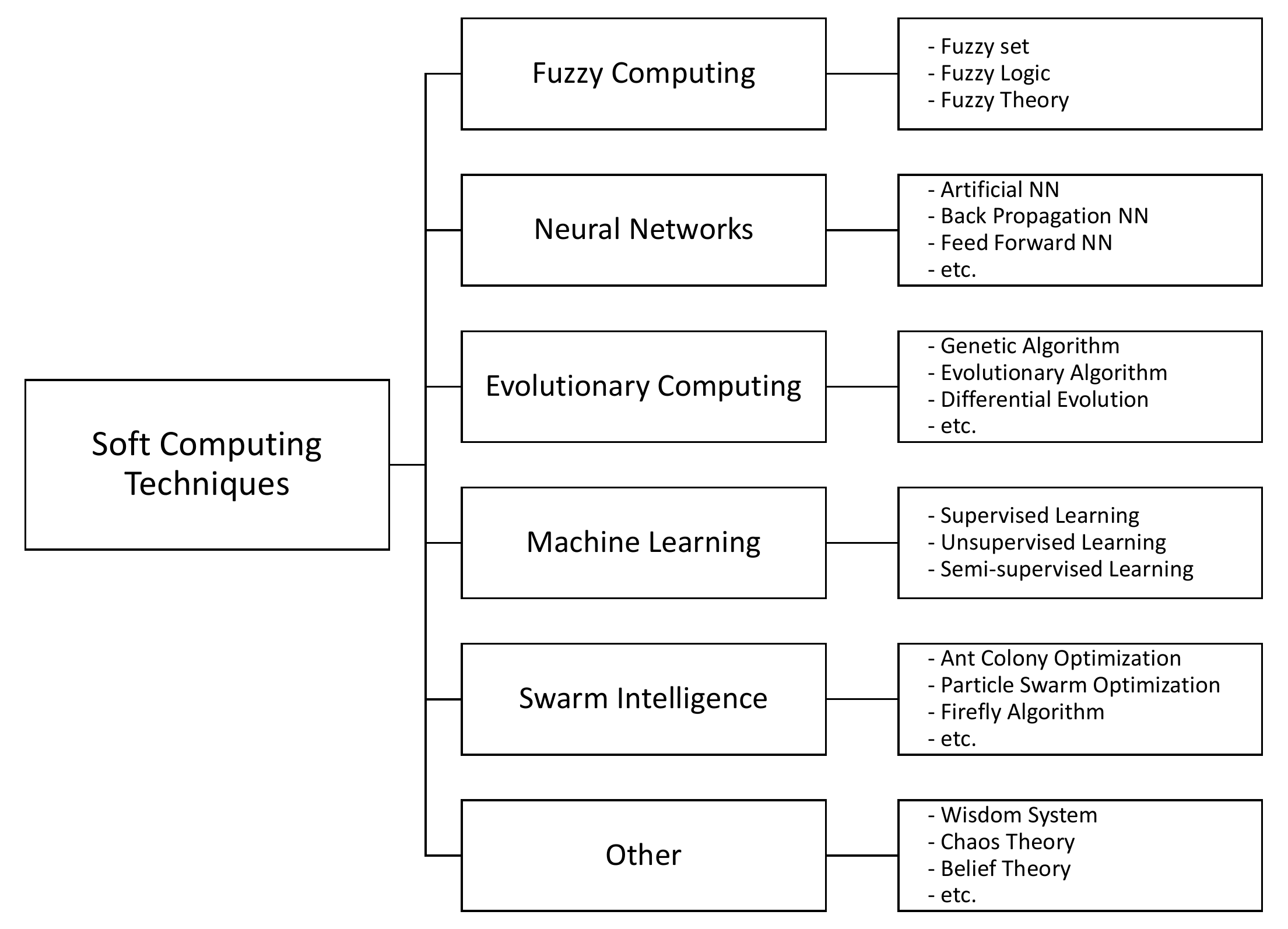}
\caption{Categories of soft computing techniques}%
\label{fig:soft_computing}
\end{figure}

As data is becoming the central point of discoveries in many scientific fields, the need for synthesis of different sources of knowledge led to challenges related to data imprecision and uncertainty. This increased the usage of soft computing techniques because of their inherent robustness and ability to operate in these difficult settings~\cite{Corona2018}. 
Soft computing, as the opposite of hard computing, is based on probabilistic modelling, approximation and flexibility~\cite{CompInt2016}.
Enabling us to grapple with theoretical problems that are analytically intractable, there is a wide range of algorithms that fall under the umbrella we call soft computing: fuzzy computing, neural networks, evolutionary computing, machine learning (ML), swarm intelligence, and other (Figure~\ref{fig:soft_computing}). 

While handling with difficulties like non-linearity, instability, and a lack of sufficient and precise information about the system, they may not always succeed in predicting the correct or exact solution, but can often come up with approximate solutions that are robust.
More importantly, soft computing provides an efficient means to deal with computationally demanding problems by processing information in a reasonable amount of time, for which traditional means are incapable.

\subsection{Sustainability Science}

Although the term \textit{sustainability} is a hot topic in various research and government domains, there is a certain degree of ambiguity in its understanding, especially in software engineering. 
Without a clear and commonly accepted definition, the research endeavours in sustainability science remain isolated and limited, leading to ineffective conclusions~\cite{Penzenstadler2013}.
A general agreement was given by United Nations Brundtland Commission in 1987, when sustainability was  defined as “\textit{meeting the needs of the present without compromising the ability of future generations to meet their own needs}”~\cite{Thomsen2013}.

Among the software quality attributes, as defined by the ISO/IEC 9126 standard~\cite{ISO9126}, the attribute of efficiency and its sub-attributes of time behaviour, resource utilization and efficiency compliance are mainly regarded as sustainability-related ones.
Efficiency is defined by the relationship between the level of performance of the software and the amount of resources it uses, revealing a very important trade-off in soft computing. 
Algorithmic efficiency in soft computing applications is not as straightforward as in traditional problems which have a clearer notion of task difficulty, like sorting for example, where the “difficulty” of the problem is the length of the list~\cite{goh2021multimodal}.
However, a similar \textit{efficiency lens} may be applied to soft computing by finding ways to reduce the required amount of time or energy and, simultaneously, to maintain the performance level.

\subsection{The Importance of Data Pre-processing}

Modern soft computing applications are often based on raw data taken from the source, which suffers from inconsistencies and errors or simply is not suitable for a data mining process~\cite{alexandropoulos2019}. 
It is the task of data pre-processing to integrate data from various sources, clean it, denoise, normalize and transform it to satisfy the requirements of soft computing algorithms~\cite{garcia2015data}.
This phase is also very important from the perspective of algorithmic efficiency, because activities like data cleaning and feature selection or extraction may reduce the amount and dimensionality of data that is repeatedly used to train, test and fine-tune the models~\cite{schneider2019principles}.
The analysis of most popular ML algorithms such as random forests (RF), decision trees (DT), artificial neural networks (ANN), and support vector machines (SVM) for analyzing biomedical data demonstrated the benefit of using information gain-based dimension reduction on models' efficiency~\cite{Deng2021}. It allows for a reduction of training time by one order of magnitude, while maintaining the overall accuracy of the prediction model.


Data cleaning is one of the simplest tasks that is often neglected or disorganized in the data mining process, resulting in out-of-range values, impossible data combinations, missing or inconsistent values, redundant information and, consequently, deteriorating the performance of trained models and the validity of studies~\cite{huebner2016systematic,lakshmi2018overview}.


Feature selection is used to remove strongly correlated features and features which are not related to class variables, but it also enables to train more robust classifiers than the ones using the original large feature space~\cite{khaire2019stability}.
Two main categories of approaches are wrapper methods, which repeatedly evaluate the model performance after selecting different subsets of features and filter methods, which estimate the features by computing their inherent properties.
Filter methods only rank the features, leaving the final choice to data scientist, and yield the results much more efficiently because they use simpler techniques like: 
\begin{itemize}
    \item Chi-squared - looking for (in)dependent features,
    \item ANOVA - looking for significantly different features,
    \item Mutual information - looking for maximal information gain,
    \item Kendall’s correlation - looking for difference in order, i.e. association strength, 
    \item Pearson or Spearman correlation, depending whether the data satisfy the requirements of normality, continuous values and small difference in variances.
\end{itemize}

Feature extraction is another mechanism of reducing the dimensionality of data, but instead of omitting features, it combines them.
Two main categories of approaches are linear methods like principal component analysis (PCA), linear component analysis, factor analysis or multi-omics factor analysis and non-Linear methods like autoencoders, representation learning or t-distributed stochastic neighbor embedding~\cite{mirza2019machine}.
The most popular method PCA is based on covariance analysis of features and it creates equal number of new components that are ortogonal, i.e. linearly independent. 
The components act as meta-features whose rank indicates their importance, with the first one containing the highest amount of variance within the dataset.
The final choice of components is again left to data scientist, but with a clear notion of cumulative variance they represent and with guidelines like the Kaiser criterion that suggests to keep components whose eigenvalue is greater than 1~\cite{kaiser1960application}.

\section{Training Course}
The aim of this training is to introduce sustainability from the perspective of soft computing, namely ML, which presents an important element for the software-driven future. 
With the principles of green data mining in focus, the significance of using data pre-processing to improve the algorithmic efficiency of training predictive models will be addressed.
After finishing this course, the key messages to take away will be:
\begin{enumerate}
    \item The importance of data pre-processing in modern soft computing applications; 
    \item The benefits of using feature selection and extraction;
    \item The guidelines on how to develop your own sustainable ML framework.
\end{enumerate}


\subsection{Prerequisite Knowledge and Skills}

The intended audience of this lecture are students of master or doctoral study programme of computing, software engineering or related programmes.
For the participants to be able to follow the training materials, basic understanding of ML and the following concepts are recommended, but not mandatory:
\begin{itemize}
    \item The difference between regression, classification and clustering, i.e. supervised and unsupervised learning;
    \item The principle of training and validating a ML model;
    \item Evaluation metrics for estimating the performance of a prediction model;
    \item The concept of multivariate analysis. 
\end{itemize}

\subsection{Materials and Methods}

Running Average Power Limit (RAPL) provides a set of performance monitoring counters to estimate the energy consumption for CPUs and DRAM~\cite{RAPL2019}. 
Being a software power model, it does not collect analog measurements, but instead makes periodical readings (100 $\mu s$ to 1 $ms$) of performance counters and I/O models~\cite{energy2015}. 
Its prediction matches actual power measurements, yields negligible performance overhead~\cite{khan2018rapl} and thus presents a promising tool to measure and monitor the energy consumption.

To demonstrate how feature selection and extraction can improve the efficiency of ML-based software systems, the accuracy and total energy consumption will be measured and compared for the following three scenarios:
\begin{enumerate}
    \item Training the ML model with all available features. 
    \item Performing feature selection and training the model on the chosen subset. 
    \item Performing feature extraction and training the model on the constructed meta-features. 
\end{enumerate}

The UC Irvine (UCI) repository\footnote{https://archive.ics.uci.edu/ml/index.php}, which hosts 622 data sets as a service to the ML community, is used in numerous empirical analyses and currently (in January 2022) it has over 7000 citations according to Google Scholar~\cite{UCI}. It provides a platform of databases, domain theories, and data generators and will be used for this tutorial as a source of data from various application domains.
The packages we plan on utilizing for the computational part of the tutorial are:
\begin{itemize}
    \item Keras - model construction and training;
    \item pyRAPL - energy consumption measurement;
    \item Scikit-learn - performing feature selection;
    \item Matplotlib - visualizing the results;
    \item NumPy - supporting mathematical operations;
    \item Pandas - manipulating data.
\end{itemize}

\section*{Acknowledgement}

This paper acknowledges the support of the Erasmus+ Key Action 2 (Strategic partnership for higher education) project No. 2020-1-PT01-KA203-078646: “SusTrainable - Promoting Sustainability as a Fundamental Driver in Software Development Training and Education”.

The information and views set out in this paper are those of the author(s) and do not necessarily reflect the official opinion of the European Union. Neither the European Union institutions and bodies nor any person acting on their behalf may be held responsible for the use which may be made of the information contained therein.

\title{Energy-driven software engineering}
\subtitle{Short lecture notes}
\author{Ana Oprescu\inst{1}\orcidID{} \and
Lukas Koedijk\inst{2}\orcidID{} \and
Sander van Oostveen\inst{3}\orcidID{}\and
Stephan Kok\inst{2}\orcidID{}}
\authorrunning{A. Oprescu et al.}
\institute{Complex Cyber Infrastructure, University of Amsterdam, The Netherlands \email{a.m.oprescu@uva.nl} \and
KPMG
Software Engineering Master, University of Amsterdam, The Netherlands\\
\and
Informatics Institute, University of Amsterdam, The Netherlands\\
}
\maketitle
\begin{abstract}
The EU Climate Law targets climate neutrality by 2050. At the same time, the Knowledge Economy is deemed crucial to EU prosperity.  Thus, the energy footprint of software services remains an important topic of research and education. In these lecture notes, we set out to educate students on energy-driven software engineering. We show how to establish whether some programming languages are inherently more “green”. We also show how to assess whether established software smells (anti-patterns) are also energy-related smells. 
Possible directions include investigating the impact of the difference at code level on the energy consumption. 

\end{abstract}
%
%
%

\section{Introduction}
Nowadays it seems like more and more people are concerned with global warming. Global warming is partly the result of the emission of greenhouse gasses during energy generation of conventional energy options~\cite{schneider1989greenhouse}. One solution for this problem is to change to green energy generation. Another solution is to decrease the energy consumption, which is not only good for the environment, but can also save a lot of money on the energy bill. 

The energy consumption of communication networks, personal computers and data centers worldwide is increasing every year~\cite{van2014trends}. This happens at a growth rate of $10\%$, $5\%$ and $4\%$ respectively~\cite{van2014trends}, therefore it is important to research ways of decreasing the energy consumption. In the field of hardware there is, according to Koomey's law~\cite{koomey2010implications}, an increase in the number of computations per Joule. However this is not enough, because tasks are designed to need more computations to complete due to the confidence in the improvement of hardware~\cite{verdecchia2017estimating}. For this reason we need to look at possibilities in decreasing the energy consumption from a software perspective.

The software perspective can include a diverse range of views, such as developers awareness, knowledge about software energy consumption, tooling problems, promoting awareness, production behaviour and maintenance requirements~\cite{Ournani2020}. In these lecture notes, we focus on programming languages (tooling) choices.

\subsection{Contents of the intended lecture}
Green thinking in software development is a key aspect of meeting the sustainability goals of the European Union, especially in the context of a Knowledge Economy which relies on digital services. Many academic level educational programs are in need of developing a ‘Green thinking’ theme~\cite{Torre2017}.

The ‘Green Thinking’ dimension should be introduced as one of the default educational requirements. The dimension should not be graded in terms of how ‘green’ the final product is, rather on how in-depth the analysis of the development options in terms of sustainability is~\cite{Saraiva2021}.

\subsection{Learning Objectives}
To initiate the integration of `Green Thinking' in modern computer science curricula, we formulate the following learning objectives:
\begin{itemize}
    \item Programming languages do have different energy footprints
    \item Energy smells do exist and have different impact depending on the programming language under observation
    \item Code level analysis can yield different energy footprints for different types of statements
\end{itemize}

\subsection{Energy footprints of programming languages}
Based on research conducted by Koedijk~\cite{LukasMSc}, we first introduce students to the concept of `Green Thinking' by exploring together two aspects of energy-aware choices regarding programming languages. Ideally, this first activity should familiarize the students with the methodology to determine whether it is the programming language or the application that dictates the magnitude of energy consumption; of course, they could both be equally dominant. 

\begin{itemize}
    \item Is there a difference in the energy consumption of software projects in different programming languages that have the same functionality?
    \item Is there a difference in the energy consumption of different software projects (using the same programming language) that have the same functionality?
\end{itemize}

\subsection{Energy smells: code smells with respect to energy consumption}
Code smells are established as undesirable code patterns identified several decades ago by Martin Fowler~\cite{fowler1997refactoring}, with a focus on readability and extensibility. Meanwhile, the term has become generally applicable to undesirable behaviour, and we apply it to code patterns that would lead to a more energy hungry behaviour as `energy smells'.

Based on research conducted by Kok~\cite{KokMSc} and Oostveen~\cite{OostveenBSc}, we introduce the students to two approaches to `energy smells' detection: \textit{refactoring} established undesirable code patterns and \textit{inflicting} established undesirable code patterns. 

\begin{itemize}
    \item What is the impact of refactoring code smells on the energy consumption of Java based open-source software projects?
    \begin{itemize}
        \item Long Function, Large Class, Duplicate Code~\cite{fowler1997refactoring}
    \end{itemize}
    \item Is the impact of a code smell on the energy consumption of a software project different between computer languages?
\end{itemize}

\section{Energy measurement approaches}
Table~\ref{tab:en-approaches} summarises the energy measurement approaches that we plan to use in our educational activities. We will use both hardware and software energy measuring approaches, to understand the sensitivity of the problem, as software approaches are coarser grained, but more accessible than hardware approaches. We will start with a software approach, such as RAPL-based measurement libraries~\cite{rapl-desrochers2016validation}. We will continue with a hardware approach, to understand how to gauge the sensitivity of an application to the measurement granularity. To that end, we will leverage dedicated state-of-the-art equipment available in DAS.
 
\begin{table}
\caption{Energy measurement approaches\label{tab:en-approaches}}
\begin{tabular}{p{.6\linewidth}|c|c}
~ & Current system power & Cumulative \\
\hline
  \textbf{Hardware}: accurate, expensive set-up & &  \\ 
\url{https://github.com/lukaskoedijk/Green-Software} & &  \\ 
a Ractivity PDU based framework  & \ding{51} & \ding{51} 
\\\hline 
  \textbf{Software}: less accurate, more accessible & & \\
\url{https://github.com/sandervano/GreenCodeSmells} & & \\ 
a RAPL-based C implementation depends on the architecture & & \ding{51}
\end{tabular}
\end{table}




\section{Energy footprints of programming languages}
To investigate the energy footprints of programming languages we selected the following programming languages: Java, JavaScript, Python, PHP, Ruby, C, C++ and C\#. We will target software projects representing several problems, each problem implemented in each programming language in several ways. We will use the Computer Language Benchmarks Game\footnote{\url{https://benchmarksgame-team.pages.debian.net/benchmarksgame/index.html}} to retrieve implementations of such problems in the selected programming languages. Originally, Koedijk has used the current system power measurement using specialized hardware\footnote{\url{https://github.com/lukaskoedijk/Green-Software}} to perform the energy measurements. We plan to use here both a software approach (RAPL-based) and the original hardware approach. 

We compare two programming languages by statistically testing the difference in energy consumption between programs solving the same problem in each programming language. We use twice the one-sided Mann Whitney U test~\cite{nachar2008mann} on the measurements of the programs. We calculate this for every language with every other language and create a table for every problem. Table~\ref{tab:lang-example} shows the comparison of different languages for the NBody problem. In such a table, a \textit{+} means that the programming language on the row is performing better than the programming language on the column, i.e.\ the programming language on the row consumes less energy. The \textit{-} means the opposite, a \textit{0} means equal and \textit{unknown} means that both one-sided Mann Whitney U tests could not be rejected. 

\begin{table}[h]
\centering
\resizebox{\textwidth}{!}{%
\begin{tabular}{|l|c|c|c|c|c|c|c|c|c|c|}
\hline
 & \multicolumn{1}{l|}{Java} & \multicolumn{1}{l|}{JavaScript} & \multicolumn{1}{l|}{Python} & \multicolumn{1}{l|}{PHP} & \multicolumn{1}{l|}{C\#} & \multicolumn{1}{l|}{Ruby} & \multicolumn{1}{l|}{C-flags} & \multicolumn{1}{l|}{C-noflags} & \multicolumn{1}{l|}{C++-flags} & \multicolumn{1}{l|}{C++-noflags} \\ \hline
Java& 0 & + & + & + & + & + & - & + & - & +\\ \hline
JavaScript & - & 0 & + & + & + & + & - & + & - & +\\ \hline
Python& - & - & 0 & - & - & - & - & - & - & -\\ \hline
PHP & - & - & + & 0 & - & - & - & - & - & -\\ \hline
C\# & - & - & + & + & 0 & + & - & + & - & +\\ \hline
Ruby & - & - & + & + & - & 0 & - & - & - & -\\ \hline
C-flags & + & + & + & + & + & + & 0 & + & Unknown & +\\ \hline
C-noflags & - & - & + & + & - & + & - & 0 & - & +\\ \hline
C++-flags & + & + & + & + & + & + & Unknown & + & 0 & +\\ \hline
C++-noflags & - & - & + & + & - & + & - & - & - & 0\\ \hline
\end{tabular}%
}
\caption{The comparison of the different languages for the NBody problem on \textit{node28}. A \textit{+} means that the language on the row consumes less energy than the language on the column, the opposite for \textit{-}, and the \textit{Unknown} means that we could not reject the null hypothesis.}%
\label{tab:lang-example}
\end{table}

\begin{figure}[H]
  \centering
  \subfloat[Programming languages comparison heatmap. Green
means a lower energy consumption for the row's programming language than the column's.\label{fig:greener}]{
\includegraphics[scale=0.5]{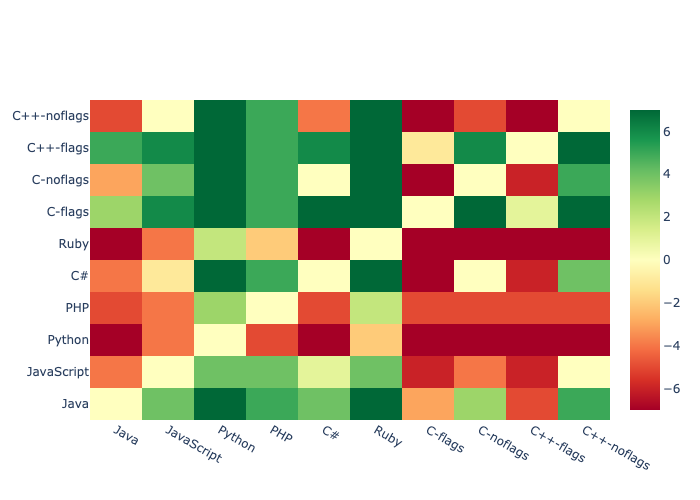}} \\
  \subfloat[Heatmap of the amount of times the null hypothesis could not be rejected.\label{fig:sometimes-not}]{
\includegraphics[scale=0.5]{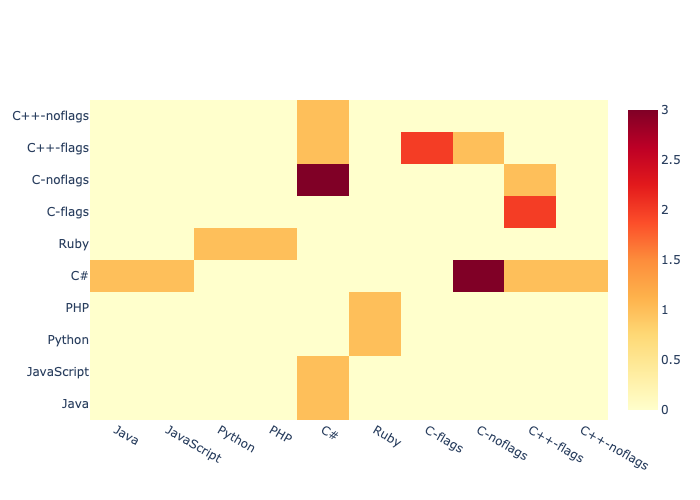}
  }
\caption{Energy consumption across multiple programming languages and applications}%
\label{fig:1}
\end{figure}

Such tables show which language is performing better compared to others for a single problem at a time. To give a total overview of problems and programming languages we calculate a score for every combination. One point is rewarded when there is a plus, one point is subtracted when there is a minus and nothing is added nor subtracted in the case of a zero or unknown. With the use of these scores we create a heatmap, such as in Figure~\ref{fig:greener}, where green means a high score (outperforming) and red a low score (underperforming).
We also check for which programming language combination \textit{unknown} occurred the most, and create a heatmap where a one is added for every \textit{unknown}, such as in Figure~\ref{fig:sometimes-not}.

\subsection{Findings from the original study}
Comparing the programming languages to each other showed that C-flags and C++-flags perform the best regarding the energy consumption for every problem. We also expect the compilation flags to play a part in the energy consumption of C and C++. The results indicate that the programs written in C and C++ compiled with compilation flags perform better regarding the energy consumption than those obtained without compilation flags.

Because there is a difference between programming languages that have a pre-compilation phase and those that do not, we also analyzed the programming languages without a pre-compilation phase separately. JavaScript performs better than the other programming languages without a pre-compilation phase.

Pereira et al.\ also analysed the energy consumption of programming languages~\cite{pereira2017energy}. They find that the programming language C consumes the least amount of energy overall and that a faster program does not imply less energy consumed, matching our findings.

When comparing two almost identical programs, we find that a for-loop uses less energy than a while-loop. We also compare the difference between having the body and condition of an if-statement on the same or different lines and find that the placement of the body might not matter.

\subsection{Educational activities}
Source code level analysis has indicated that a for-loop uses less energy than a while-loop and that it does not matter if the body of an if-statement is on the same or on a different line as the condition. We plan to focus on replicating the methodology for the ternary operator as an educational activity.

\section{Energy smells}
A promising approach to find energy hungry code patterns is to investigate whether code smells related to software quality are also energy smells~\cite{KokMSc,OostveenBSc}.

\subsection{Refactoring energy smells}
Kok tries to assess the energy footprint of code smells by measuring the impact of established refactorings for the selected code smells on the energy consumption~\cite{KokMSc}. 

We measure the impact of refactorings in terms of software metrics via BetterCodeHub\footnote{\url{https://bettercodehub.com/}}. We select the Java programming language and the JDeodorant code smell detection tool~\cite{tsantalis2018ten}. We measure the energy consumption of code executed via the JUnit test suite. We selected software projects have solid testing framework with high code coverage, open source and medium sized: \texttt{FasterXML/java-classmate}, \texttt{apache/commons-lang},\\\texttt{apache/commons-configuration}. The code smells under investigation are \texttt{`Long Function'}, \texttt{`Large Class'}, and \texttt{`Duplicate Code'}~\cite{fowler1997refactoring}. We plan to use here both a software approach (RAPL-based) and the original hardware approach\footnote{current system power measurement using specialized hardware (\url{https://github.com/lukaskoedijk/Green-Software})}.

\subsubsection{Findings from the original study}
There is an increase in energy consumption after refactoring for `Long Function' and `Large Class'. Additionally, when refactoring for `Duplicate Code' there is both an increase and decrease in energy consumption. In a nutshell, developers can safely write quality code while at most negligibly increasing the energy consumption of the code base; sometimes it might even yield a slight decrease in the energy consumption.

\subsection{Inflicting energy smells}
Oostveen takes another path to assess the overlap of energy smells and traditional code smells, namely introducing the latter in a controlled manner in existing code bases~\cite{OostveenBSc}. 

We select the following programming languages: Java, Python, C++. In terms of software projects, we select several problems, each problem implemented in each programming language in several ways. We choose the problems from the Computer Language Benchmarks Game\footnote{\url{https://benchmarksgame-team.pages.debian.net/benchmarksgame/index.html}}: Fasta (memory intensive) and NBody (CPU intensive), and from Project Euler\footnote{\url{https://projecteuler.net}}: Sudoku with a DFS approach (CPU intensive). We introduce in each project the same code smell, namely \texttt{‘Long Function’}. We measure impact of introduced code smells in terms of software metrics via BetterCodeHub. We plan to use here both the original study cumulative energy measurement using internal software\footnote{\url{https://github.com/sandervano/GreenCodeSmells}} and a hardware approach.


\subsubsection{Findings from the original study} 
In general, the presence of the `Long Function' code smell increased the energy consumption of software projects written in different computer languages. The impact of the `Long Function' code smell on the energy consumption of a software project is different between different computer languages. The impact of the `Long Function' code smell on the energy consumption of a software project is different between different problems written in the same computer language.

\subsection{Educational activities}
Inflicting established code smells in a controlled manner yielded interesting results. We plan to apply this approach to other established code smells, such as `Large Class' and `Duplicate Code'.

\title{Using Strong Types in \CPPTitle{} for Long-term Code Management
    \thanks{Prepared with the professional support of the Doctoral Student Scholarship Programme of the Co-operative Doctoral Programme of the Ministry of Innovation and Technology, financed from the National Research, Development, and Innovation Fund.}%
    }
\titlerunning{Using Strong Types in \CC{} for Long-term Code Management}
\author{Richárd Szalay \orcidID{0000-0001-5684-5158} \and
    Zoltán Porkoláb \orcidID{0000-0001-6819-0224}}
\authorrunning{R. Szalay and Z. Porkoláb}
\institute{Department of Programming Languages and Compilers, \\
    Institute of Computer Science, Faculty of Informatics, \\
    ELTE Eötvös Loránd University, Budapest, Hungary \\
    \email{szalayrichard@inf.elte.hu}, \email{gsd@ik.elte.hu}%
    }
\maketitle
\begin{abstract}
When using programming languages, type systems are crucial tools in the hands of developers to guarantee an elevated level of safety of their programs.
However type systems are not used to their full extent in practice, and trade-offs are made.
This mainly manifests in developers overusing built-in and well-known library types, such as \mintinline{CPP}{int} or \mintinline{CSharp}{string}, instead of types that express the domain of the represented values with a finer granularity.
The results of this range from a hindrance to code comprehension to subtle security vulnerabilities going undetected for potentially years.
Education is a crucial element in combating the trend of quality deterioration in software projects.
By teaching the new generation of developers the importance of stronger and safer types, we hope to ensure more sustainable development processes in the future.

In this paper, we detail how the education of strong typing was introduced in our \emph{Advanced \CC{}} M.Sc.\ subject.
While certain ideas of strong typing are actionable from the very early subjects in education, the hands-on experience with improving an existing, hard-to-understand code ensures students internalise the ideas better.

\end{abstract}
\section{Introduction}\label{sec:introduction}
Several large and mainstream programming languages, including \CC{}, exhibit a statically checked and strongly typed type system.
This means that the compiler checks whether an expression or instruction is allowed to be evaluated during compilation.
Developers and architects can thus use the type system to guarantee a degree of safety of their program.
\CC{} comes with a small set of \emph{fundamental types}  defined in the standard~\cite{ISO:2017:IV} to be supplied by the compiler -- as opposed to being written as code in a library.
These fundamental types include, most notably, integer (\mintinline{CPP}{int}) and floating-point (\mintinline{CPP}{float}) numerals, the \mintinline{CPP}{char} (character) type, and trivial type constructs such as pointers and arrays.
Moreover, the \emph{standard library} offers some additional generic types, such as \mintinline{CPP}{std|$::$|}\mintinline{CSharp}{string}.

Unfortunately, it has been observed that developers tend to overuse these fundamental and widely available types, e.g.\ by setting both the size of a machine part and the orientation of it to be a \mintinline{CPP}{double} or conflating the name of a person with their address in \mintinline{CSharp}{string}s.
With the types of program elements kept at a coarse level, developers resort to embedding semantic information in the \emph{names} of variables.
While this is visually indicative to a knowledgeable developer, and there are ways to use human-assigned names meaningfully in an automated context~\cite{Butler2010Exploring}, the mainstream languages and compilers ascertain no value to the identifier.
This had resulted in researchers observing dormant security vulnerabilities in large, industrial projects~\cite{Rice:2017:DAS:3152284.3133928}, with potentially many more mistakes only waiting to be made and discovered~\cite{SzalayJSS2021}.

The solution to the problem is using \emph{strong typing}~\cite{Madsen1990Strong,Bertrand1992Ensuring}, where the capabilities of the language and the type system are used to the full extent possible to ensure that the compiler can detect and thus prevent programmer errors.
Strong typing, however, is not a discrete binary state of the system but a comparable measure: some software solutions to the same problem can be \emph{more strongly typed} than another.

Two key insights justify the education of type safety and strong types.
First, it is known from previous literature that the quality of a software system generally deteriorates over time.
The application of refactoring, which improves quality metrics, can generally be traced back to a few developers' personal motivations for doing so~\cite{Szabados2016Internal,6671299}.
While reworking an existing software system to stronger types is a tedious task to do manually, creating automated tools as a generic solution is also non-trivial~\cite{Szalay2022Flexible}.
Existing solutions were created as specific algebra for particular domains, limiting their applicability~\cite{49386}.
Thus, the second reason is that by ensuring that new graduates are educated about strong typing, they can start their practice, whether industrial or research, by writing new code already satisfying the need for type safety.

\section{Laws \& Curricular Framework}\label{sec:law-and-curricula}
Higher education in Hungary is regulated by an associated legal framework.
The general requirements of \emph{Computer Science} (CS) education is described in the law, and universities have only some autonomy in defining the specifics of their education locally.
In accordance with the Bologna framework~\cite{BolognaProcess}, education is done in three tiers: Bachelor (B.Sc.), Master (M.Sc.) and Doctorate (Ph.D.).
CS education in Hungary is mandated by law to be 6~semesters (180~ECTS credits) of B.Sc.\ followed by 4~semesters (120~ECTS credits) of M.Sc., which is in stark contrast when compared to engineering training which instead use a 7:4~semester (210:120~ECTS) division.
The Faculty of Informatics's old CS curriculum was accredited in 2008 in which B.Sc.\ students were taught beginner \CC{} on a obligatory subject, and one of the M.Sc.\ specialisations contained an elective \emph{``Programming languages''} block in which modern \CC{} topics were taught on \emph{``Multi-paradigm programming''}.
This subject was allocated 3 ECTS credits for a weekly 2-hour lecture and individual work, followed by an exam, which was a test of various forms of questions but contained no dedicated coding exercise other than questions that can be answered with just a few lines of code.

The curriculum for CS education was refactored, refreshed, and reaccredited since, and the first year of students under the new system started their studies in 2018.
The B.Sc.\ study plan includes a single \emph{``Programming Languages''} subject, on which students are taught the fundamentals of programming language elements and design, currently through the examples of Java.
Beginner \CC{} education was resurrected as an elective subject.
The Multi-paradigm programming subject had been renamed \emph{``Advanced \CC{}''} and given an additional 2 ECTS credits --- now totalling 5 --- for a weekly 2-hour practice lesson in addition to the lecture.
While the system of multi-subject \emph{blocks} had been removed from the framework, Advanced \CC{} remains an elective subject on the M.Sc.\ curriculum.

Approximately 600 students enrol CS each year on the Bachelor level and 100-120 on the Master level~\cite{Felvi}.
Bachelor students are enrolled based on their secondary school final exam (\emph{Matura}) scores without a dedicated admission exam by the University.
Knowledge of foreign languages (including English) is not a requirement for enrollment.
To be given the Bachelor degree, the students must obtain a state-accredited certificate of language knowledge at the CEFRL B2 level from one foreign language.
Most CS students will do an English exam or already have done it.
Master students are enrolled based on an admission exam, and a relevant Bachelor degree is a requirement.
As the Advanced \CC{} subject is only announced for local, Hungarian students, the previous figures include only the local enrolments.
While the B.Sc.\ curriculum has a direct equivalent for international students, taught in English, the post-2018 M.Sc.\ curricula are different: several specialisations are announced only in English, and both international and local students may enrol on those.
Those students are out of scope for this education report, as their education is usually done in Python or Java.
They do not have a dedicated subject specifically for teaching a programming language.

Typically, the Faculty of Informatics offers CS education both full-time/day-time (lessons each day may start as early as 08:00) and part-time/night-time (lessons each day start after 16:00) education, but not in correspondence (lessons only one or two days per month) form.
Most Master courses are held in accordance with the night-time schedule, together with, and even for those officially enrolled as day-time students.
Also, the COVID-19 situation~\cite{COVID19} has forced the University to do fully remote or hybrid attendance education, depending on the circumstance.

As refactoring and type migration necessarily involve the understanding of an existing software system, \textbf{we expect students to have an intermediate command of English}, so they understand the documentation and comments of existing code, \textbf{and have absolved beginner \CC{} and Object-Oriented Programming skills}: they understand the fundamental data structures and operations, they know what is and how to write a class, how encapsulation and visibility works, and what \CC{} templates are.
However, we have no legal framework for checking whether the M.Sc.\ students have studied basic \CC{} as that is a B.Sc.\ subject, and direct subject-to-subject dependencies crossing study plan boundaries are not possible.

\section{The \emph{Advanced \CPPTitle{}} Subject}\label{sec:adv-cpp}
As discussed previously in \cref{sec:law-and-curricula}, the Advanced \CC{} subject in the new curriculum was made into a 5 ECTS credit subject, now including effort for a weekly 2-hour long practice lesson.
The lecture's material starts from the tricky parts of ``beginner \CC{}'' --- such as the pitfalls of special member functions --- and encompasses template metaprogramming, lambdas, exception safety, concurrency, and \mintinline{CPP}{std|$::$|visit}, remained fundamentally unchanged, along with the shape of the exam.
To ensure that the final grade reflects both the ``theory'' and the ``practice'' part, both examinations were calculated in the final grade with a $50\%$ weight.
In Hungary, a 5-scale grading system is used, where grade $1$ is \emph{Fail} and grade $5$ is \emph{Excellent}.
Assuming the exam (lecture) result and the practice is $100\%$ total, the boundaries between grades were set as follows: $40\%$ minimum for \emph{Pass}, then $55\%$, $70\%$, and $85\%$ and onwards for \emph{Excellent}.
Earning $20\%$p.\ (out of the $50\%$) each on both the exam and the assignment was also mandatory --- students doing a perfect exam but no assignment, still resulting in $50\%$ total, would not pass.

We have opted \textbf{not to do} \emph{synchronous (online)} lessons for the practice --- i.e.\ students need not be available and attend a lesson each week during practice hours.
(This is not true for the lecture, which was given synchronously, and a recording was made available for the students to watch later.)
The teachers kept the officially scheduled time for the practice available as on-demand, and due to the COVID-19 situation, online videoconferencing-based consultation.
The students were given, on $4$ occasions, a presentation-like or discussion lesson during practice hours, and then the work of the semester was the fulfilment of an assignment project out of two choices.
As both the fact that Advanced \CC{} has a practice lesson and the teaching of strong typing was a new addition to our material, we offered not just the strong typing project --- discussed in \cref{sec:teaching} --- but a \emph{library-writing} exercise: students were instructed to create a program library, in the spirit of the \emph{Standard Template Library}, using the tools and techniques presented on the lecture, that implements a container or data structure, for this semester, \emph{Trie} (or \emph{prefix tree})~\cite{10.1145/367390.367400}.

The Spring semester begins in early February, and the term time ends in the middle of May, after which the examination period lasts until the end of June.
The assignments were published in late February and were due in early May, giving the students between $2$ and $2.5$ months to complete the assignment.
The lecture exam was held in the first two weeks of the examination period (late May), and students were allowed to take both chances, with the teachers considering the better of the two attempts.
Students were also allowed to either hand the assignment in late at the very end of May if they missed the first deadline, or perform a resubmission based on the comments received on the initial submission.

\begin{table}
    \centering
    \caption{Breakdown on the results of the semester}\label{tbl:semester}
    \subfloat[][Student choice of submitted assignment]{%
        \centering
        \begin{tabular}{lrr}
            \toprule
            \textbf{All students}     & \textbf{$69$} & \textbf{$100.00\%$}  \\
            \midrule
            \textbf{Non-participant}  & \textbf{$31$} & \textbf{$44.93\%$}   \\
            Abandoned subject         & $27$          & $87.10\%$            \\
            No submission             & $4$           & $12.90\%$            \\
            Old curriculum            & $1$           & $3.23\%$             \\
            \midrule
            \textbf{Participant}      & \textbf{$38$} & \textbf{$55.07\%$}   \\
            \textit{Trie} (library)   & $30$          & $78.95\%$            \\
            \textit{Strong Typing}    & $7$           & $18.42\%$            \\
            Both                      & $1$           & $2.63\%$             \\
            \bottomrule
        \end{tabular}%
    }
    \hspace{6em}
    \subfloat[][Grades administered to students]{%
        \centering
        \begin{tabular}{lrr}
            \toprule
            \textbf{All students}             & \textbf{$69$} & \textbf{$100.00\%$}  \\
            \midrule
            0 (\textit{Miss/Abandon})         & $27$          & $39.13\%$            \\
            1 (\textit{Fail})                 & $5$           & $7.25\%$             \\
            2 (\textit{Pass})                 & $4$           & $5.80\%$             \\
            3 (\textit{Average})              & $11$          & $15.94\%$            \\
            4 (\textit{Good})                 & $16$          & $23.19\%$            \\
            5 (\textit{Excellent})            & $6$           & $8.70\%$             \\
            \midrule
            0 and 1 (incomplete)              & $32$          & $46.38\%$            \\
            2 to 5 (success)                  & $37$          & $53.62\%$            \\
            \bottomrule
        \end{tabular}%
    }
\end{table}

\cref{tbl:semester} details the statistics of the semester.
Unfortunately, the first two takeaways are that attrition of students is considerable ($45\%$ of the registered students did not partake in the assignment or even the lecture)~\cite{Takacs2021Applying} --- especially for elective subjects --- and even students who perform on the subject seem to take the perceived ``easier'' assignment.
$1$ student had registered on the subject while being on the older 2008 curriculum where the subject did not include a practice part, and thus this student was exempted from the assignments.
$4$ students took the exam but did not submit an assignment and thus failed the subject.
There was only $1$ student ($2.63\%$) out of the $38$ who submitted everything, but failed to acquire the bare minimum points to be given a pass.
Everyone else succeeded in obtaining a passing grade.

\section{Practical Strong Typing in \emph{Advanced \CC{}}}\label{sec:teaching}
\begin{figure}
    \begin{lrbox}{\mintedbox}
        \begin{minipage}{\textwidth}
            \begin{minted}{CPP}
                struct student {
                    std|$::$|string name;
                    std|$::$|string teacher;
                };

                int main() {
                    std|$::$|vector<student> students = load_students();
                    std|$::$|string name = read_name("Student name? ");
                }
            \end{minted}
        \end{minipage}
    \end{lrbox}
    \subfloat[a][Initial example with Standard Template Library types.]{%
        \label{lst:strong-type/basic}%
        \usebox{\mintedbox}%
    }

    \vspace{2em}

    \begin{lrbox}{\mintedbox}
        \begin{minipage}{\textwidth}
            \begin{minted}{CPP}
                struct student {
                    person_name name;
                    person_name teacher;
                };

                int main() {
                    std|$::$|vector<student> students = load_students();
                    person_name name = read_name("Student name? ");
                }
            \end{minted}
        \end{minipage}
    \end{lrbox}
    \subfloat[][Adding the concept of ``names'', which is a particular problem domain for strings.]{%
        \label{lst:strong-type/person-name}%
        \usebox{\mintedbox}%
    }

    \vspace{2em}

    \begin{lrbox}{\mintedbox}
        \begin{minipage}{\textwidth}
            \begin{minted}{CPP}
                struct student {
                    student_name name;
                    teacher_name teacher;
                };

                int main() {
                    std|$::$|vector<student> students = load_students();
                    student_name name = read_student_name("Student name? ");
                }
            \end{minted}
        \end{minipage}
    \end{lrbox}
    \subfloat[][Preventing argument selection defects~\cite{Rice:2017:DAS:3152284.3133928} by disjoint subdomains of \mintinline{CPP}{person_name}.]{%
        \label{lst:strong-type/student-name}%
        \usebox{\mintedbox}%
    }

    \vspace{2em}

    \begin{lrbox}{\mintedbox}
        \begin{minipage}{\textwidth}
            \begin{minted}{CPP}
                struct student {
                    student_student_name_t name;
                    student_teacher_name_t teacher;
                };

                int main() {
                    load_students_return_collection_t students = load_students();
                    read_student_name_return_t name = read_student_name("Student name? ");
                }
            \end{minted}
        \end{minipage}
    \end{lrbox}
    \subfloat[][An extremely separated version which is considered ``too strongly typed''.]{%
        \label{lst:strong-type/extreme}%
        \usebox{\mintedbox}%
    }

    \caption{Iterating a process of increasing type granularity and separation strength on a small example.}\label{lst:strong-type}
\end{figure}

During the first three weeks of the semester, students were given talks and held an oriented discussion in the time slot of the practice lesson.
This involved showing them an overview presentation about strong typing from which the keystones of the main example is depicted in \cref{lst:strong-type}.
While the semester-long subject had ample time to fit the presentation in, a day-long or half day-long training may only provide a \textbf{shortened version of such a presentation} to the students.
An example of the code snippets the students were shown is depicted in \cref{lst:strong-type}.
This example shows the different granularity of strong typing in a small code.

For the assignment, students were instructed to find an open-source and freely editable \CC{} project on the Internet that is at least ``medium sized'', identify a weak type used in it, and fix it, carrying the fix through as much of the project as possible.
Carrying the fix through meant that once the new, stronger type was introduced, it should also be given a useful encapsulated interface and existing program elements' types changed to use the new type, too.
The requirements were not strict for this, as we left it up to the students' judgement to select a project where they feel their improvements would meaningfully benefit the community as a whole.
The initial suggestion from the teachers was \emph{Doxygen}~\cite{Doxygen}, but we were up to accepting different projects.
All $8$ students chose Doxygen, however.

Students were also instructed to write some sort of a logbook or diary, as we were interested in not just the end result of the changes but the thought process during the exploration of the project.
As most active community-driven software projects nowadays use Git, creating a fork of the repository and committing with meaningful messages was often a trivial way of achieving a logbook.
The insights from their commit history were enhanced with the asynchronous discussion between the student and teachers during the evaluation of the assignment.
Students were also suggested to use the \emph{CodeCompass}~\cite{10.1145/3196321.3196352,10.1145/3196321.3197546} code comprehension tool to navigate the project.
The University hosted the server with the parse of the latest Doxygen tagged release at that time available for exploration.
Unfortunately, \textbf{these requirements}, especially the need of understanding a larger project, result in an increased workload that is \textbf{not feasible for a shorter teaching session}.
Instead, students need to be given practical \textbf{knowledge through proctored examples}.

In total, $7$ students chose and succeeded with the assignment.
$1$ student's submitted Git repository fork was impossible to understand as it contained formatting commits and trash interleaved with the supposed changes.
Coincidentally, this student was the same who chose \emph{both assignments}, and as such, they obtained their passing grade from the ``easier'' library writing instead.

\subsection{Assignments handed in by students}\label{sec:student-works}
Doxygen is a self-contained tool that deals with creating HTML documentation with graphs (in Dot or SVG formats) from source code.
To achieve this purpose, it deals primarily with strings, with the occasional numeric values for orientation, size, etc.
For numeric values, several strong typing and dimensionality libraries exist~\cite{Pusz2019Implementing}.
$4$ out of the $7$ students chose a strong wrapper type over a numeric value and created associated interfaces, e.g.\ ``width $*$ height $=$ area''.
\textbf{Such examples are good for the introduction of strong typing, even in a hand-engineered sterile context}.

The rest of the students, however, modified strings to stronger types.
While these examples may be harder at first glance, they are more rewarding as it involves changing how the \emph{values} of the type are handled in the program.
These examples will be detailed in the following.

\subsubsection{Hard-coded strings to \mintinline{CPP}{enum}}\label{sec:student/enum}
\begin{figure}
    \centering
    \includegraphics[width=\textwidth, keepaspectratio]{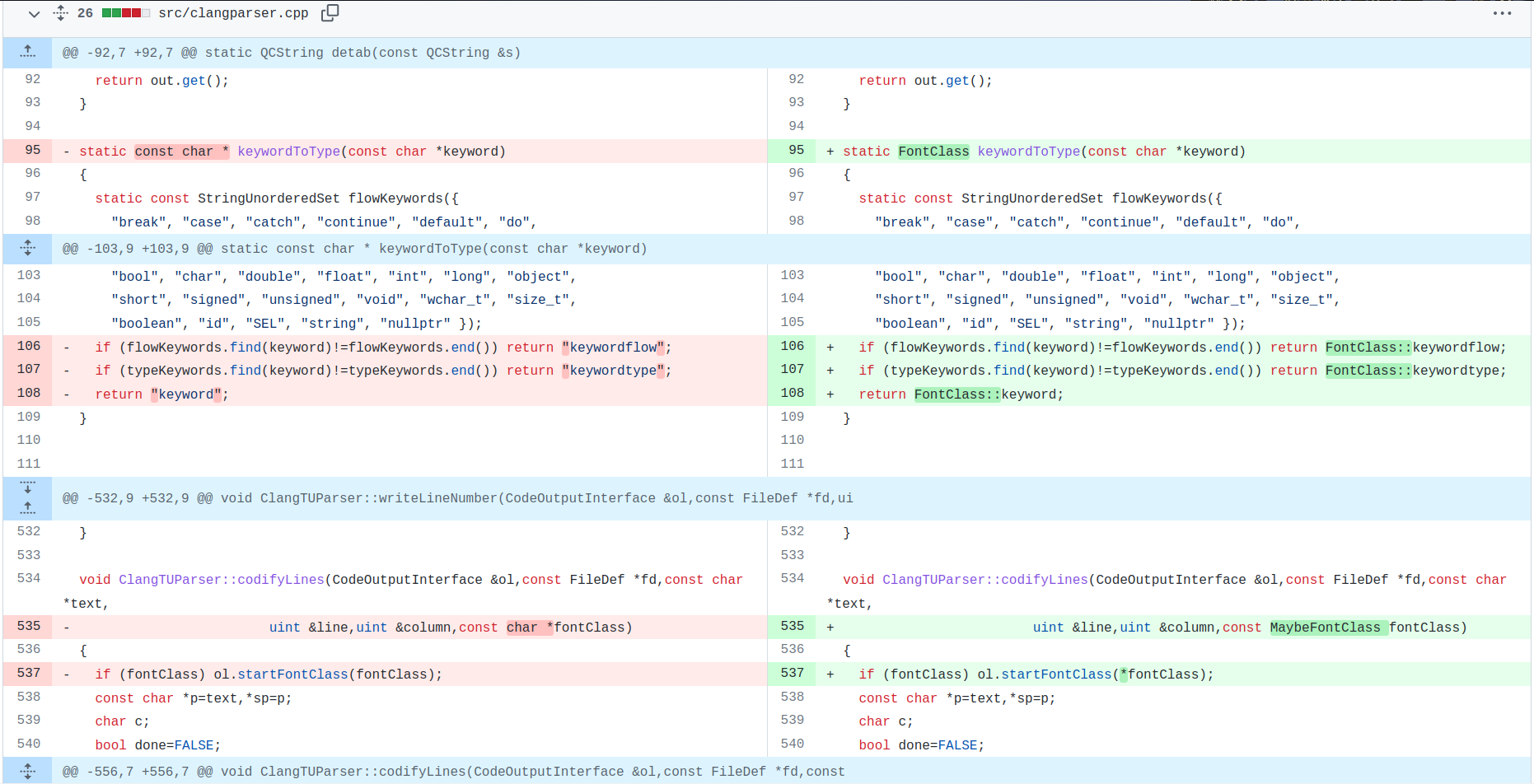}
    \caption{Difference excerpt from the assignment where hard-coded string literals were replaced with the members of an \mintinline{CPP}{enum}.}\label{fig:fontclass}
\end{figure}

Doxygen's documentation rendering uses CSS classes to highlight tokens of the source code differently.
This is handled by the \mintinline{CPP}{FontClass} value, but in upstream Doxygen, it was only passed as a C string (\mintinline{C}{const char*}) only.
As all potential values reaching the eventual printer logic was from a set of hard-coded values, the student introduced an \mintinline{C}{enum} with these values and changed the usage points accordingly.
In several places, they also added \mintinline{C}{std|$::$|optional} to represent the lack of information which was previously represented by the ``null pointer'' value~\cite{SzalayICAI2020}.
This allowed the decoupling of \emph{``we cannot determine the font-class''} from \emph{``the font class is nothing special''}.
Moreover, there are several advantages to \mintinline{C}{enum}s: using an undefined value or not handling a defined value is detected by the compiler (e.g.\ via \mintinline{Text}{-Wswitch}); and it is easier to compare as the values are well-defined integer constants, not pointers to some potentially changing memory address -- which is how string literals are represented at run-time.
\cref{fig:fontclass} shows an example, rendered as a code difference, from the student's submission.
\textbf{This is a beginner level example that highlights the benefits of an otherwise seldom-used language element.}

\subsubsection{URLs}
\begin{figure}
    \centering
    \includegraphics[width=.73\textwidth, keepaspectratio]{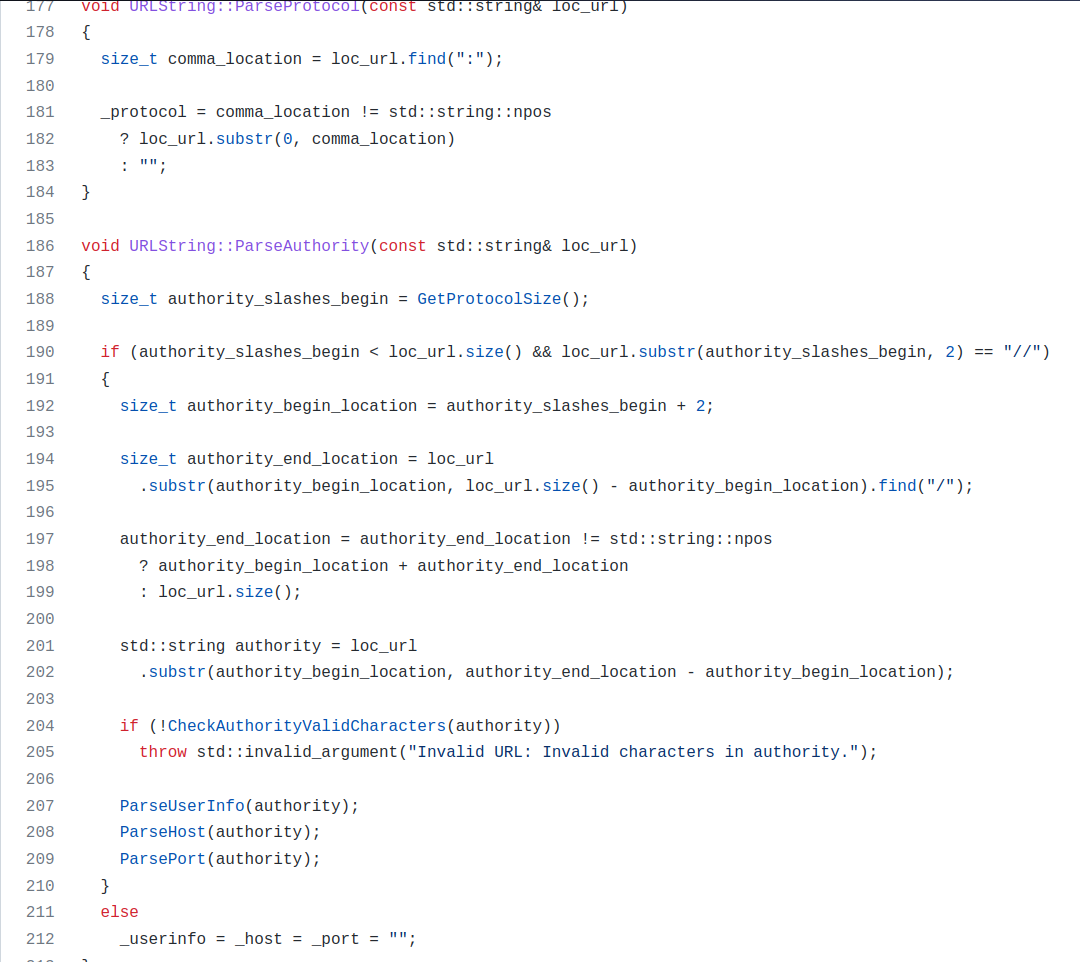}
    \caption{Implementing custom parsing logic and domain-specific error detection for representing URLs.}\label{fig:urlstring}
\end{figure}

As the output of Doxygen is a website, the system naturally also has to deal with a special subdomain of strings, URLs.
URLs are structured data represented in textual format in a standardised way~\cite{rfc1738}.
\mintinline{CSharp}{String} types in almost all programming languages allow operations that are not meaningful for URLs, e.g.\ consider a call that, potentially by mistake, boils down to \linebreak \mintinline{Python}{"http://example.com".substr(3, 7)}.
The result of this operation is \mintinline{Python}{"p://e"}, something that is not useful when observing and invalid in the problem domain of URLs when manipulating.
The student implemented a strong type, \mintinline{CPP}{URLString}, which can be constructed from a string and performs parsing of the constituents --- e.g.\ the host name, the path, or the protocol --- of the URL.
Afterwards, client code might manipulate the individual constituents.
When needed as a plain string, the instance can convert back to a single string value via calling the \mintinline{CPP}{operator()}.
\textbf{This is a complex example, as the developers need to actively deeply investigate the usage of the value, and add problem-specific logic into the implementation}; however, this is the example in which ``true'' strong typing was created.

\section{Conclusion}\label{sec:conclusion}
Type systems and the practice of strong typing are powerful tools to increase the guarantee that developer mistakes are caught in time before causing serious damage.
As large systems evolve, they tend to erode away their internal quality.
Combating entropy is usually attributed to a few developers per project for whom it is a personal, internalised goal to do so.
For this reason, it is important that the next generation of practitioners and researchers are proactively taught the importance of type safety from the very start.
By incorporating the education of strong typing into the material of advanced programming subjects, we ensure that students have the chance to expose themselves to actual, existing and widely used bad code and to obtain hands-on experience in fixing it.
This also grows their ability to navigate and understand codebases created by others, something that is also a valuable skill for life in this field.

\section*{Acknowledgements}
This paper acknowledges the support of the Erasmus\nolinebreak\hspace{-.05em}\raisebox{.4ex}{\tiny\bf +} Key Action~2 (\emph{Strategic partnership for higher education}) project \textnumero{} \texttt{2020-1-PT01-KA203-078646}: \emph{``SusTrainable -- Promoting Sustainability as a Fundamental Driver in Software Development Training and Education''}.
The information and views set out in this paper are those of the authors and do not necessarily reflect the official opinion of the European Union.
Neither European Union institutions and bodies, nor any person acting on their behalf may be held responsible for the use which may be made of the information contained therein.

\title{Sustainable Distributed Communication Models}
\author{Jianhao Li \and Vikt\'oria Zs\'ok}
\authorrunning{J. Li \and V. Zs\'ok }
\institute{E\"otv\"os Lor\'and University, Faculty of Informatics\\
Department of Programming Languages and Compilers\\
H-1117 Budapest, P\'azm\'any P\'eter s\'et\'any 1/C., Hungary\\
\email{lijianhao288@hotmail.com, zsv@inf.elte.hu}\\
}
\maketitle
\lstdefinestyle{numbers}{numbers=right, stepnumber=1, numberstyle=\tiny, numbersep=-5pt}

\lstset{language=Go,style=numbers,frame=single,
  basicstyle=\ttfamily\scriptsize,
  keywordstyle=\color{black}\ttfamily,
  commentstyle=\ttfamily\color{black},
  stringstyle=\ttfamily\color{black}}
 
\lstnewenvironment{Go}{\lstset{language=Go,frame=single,
  basicstyle=\ttfamily\scriptsize,
  keywordstyle=\color{black}\ttfamily,
  stringstyle=\color{black}\ttfamily,
  commentstyle=\color{black}\ttfamily}}{}

\begin{abstract}
In this tutorial, we analyze the existing distributed communication models of \textsc{RabbitMQ} and \textsc{Mangos}. The goal is to investigate them from scalability and usability perspectives.
Practical examples are designed to illustrate their features. Additionally, we introduce a new sustainable distributed communication model and its implementation.
The motivation is to design and implement a new distributed communication tool with outstanding usability and scalability.

\end{abstract}

\section{Introduction}
The topic of this tutorial is sustainable distributed communication models. The goal is to analyze the existing distributed communication models and their implementations. Therefore, we developed practical examples that can be easily understood when teaching. Additionally, we design and implement new distributed communication models that are more performant. 
This tutorial is meant for students who are interested in distributed communications and their implementations. The knowledge of the basics of \textsc{Go}~\cite{url_Golang} and \textsc{RabbitMQ}~\cite{url_Rabbit} is assumed.

The paper first presents examples of \textsc{RabbitMQ} and \textsc{Mangos}~\cite{url_Mangos} communication protocols. Second, a comparison of their distributed features is provided.
Finally, a new distributed communication model is introduced, which is designed based on novel approaches after studying the advantages and disadvantages of \textsc{RabbitMQ} and \textsc{Mangos}. 

The idea of designing a new model was triggered when trying to transfer the previous distributed storage project~\cite{msc_thesis} using \textsc{RabbitMQ} into a peer-to-peer structure in order to study P2P algorithms. We found that the \textsc{RabbitMQ} broker is inherently centralized. Although you can increase the reliability of \textsc{RabbitMQ} by building a broker cluster with replicated data, it is still not strictly decentralized.
However, we expect that a distributed communication tool is strictly decentralized and symmetrical, i.e.\ it has no centralized services and it is composed of equal role nodes. It should also support the implementation of a structured overlay. The structured overlay is defined as ``an overlay in which nodes cooperatively maintain routing information about how to reach all nodes in the overlay''~\cite{overlay}.

When studying other nowadays distributed communication tools, some drawbacks could be observed. \textsc{NSQ}~\cite{url_NSQ} does not meet our expectations because it is not strictly symmetrical, as it uses \texttt{nsqlookup} daemon to manage the nodes' topology. \textsc{Mangos}, a distributed communication tool implemented in \textsc{Go}, also has a weakness in usability because of the strictly separated communication patterns. 
\textsc{Erlang}~\cite{url_Erlang} has powerful distributed messaging system following the actor model. However, it needs an extra virtual environment and it lacks constructs with built-in mechanisms that enable programmers to easily implement systems with complex distributed communication topologies.
Therefore, we decided to design a new distributed communication model and to implement a new tool, which is strictly decentralized and symmetrical, and its main advantage is the excellent usability.

\section{Existing distributed communication models}

In this section, we introduce two existing distributed communication models of \textsc{RabbitMQ} and \textsc{Mangos} with practical examples. At the end of the section, we compare them and draw conclusions.

\subsection{\textsc{RabbitMQ}}

\begin{figure}[!htb]
\centering
\includegraphics[width=1\textwidth]{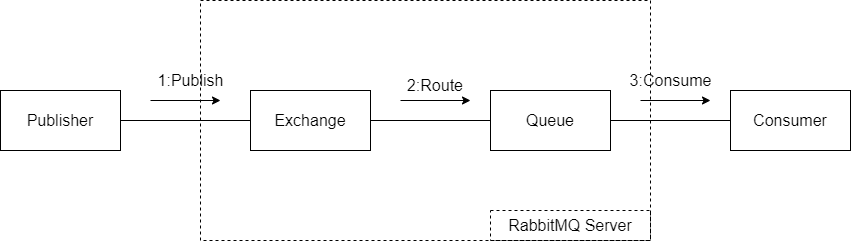}
\caption{RabbitMQ\label{fig:RabbitMQ}}
\end{figure}

\textsc{RabbitMQ} is open-source message-broker software and it implements the \texttt{AMQP} 0-9-1 protocol~\cite{url_amqp}. Its communication model is shown in~\autoref{fig:RabbitMQ}. The \texttt{publisher} and the \texttt{consumer} are \textsc{Go} programs, and they use the \textsc{RabbitMQ} client library to connect to the \textsc{RabbitMQ} server. There can be multiple publishers and consumers running on separate computers. The \texttt{publisher} publishes messages to the \texttt{exchange} of the \textsc{RabbitMQ} server with a publishing routing key. The \texttt{queue} is connected to the \texttt{exchange} with a bounding routing key. The \textsc{RabbitMQ} server routes messages from the \texttt{exchange} to the \texttt{queue} based on the type of the \texttt{exchange} and the matching routing keys. The \texttt{exchange} can not store messages, while the \texttt{queue} can.

An example of distributed messaging illustrates the mechanisms of fanout exchange, direct exchange and shared queue. The fanout exchange distributes messages to all the queues bound to it. The direct exchange routes the messages based on the publishing routing key matched against the bounding routing key.
The consumers sharing a queue receive the messages in a round-robin way.

\begin{figure}[!htb]
\centering
\includegraphics[width=1\textwidth]{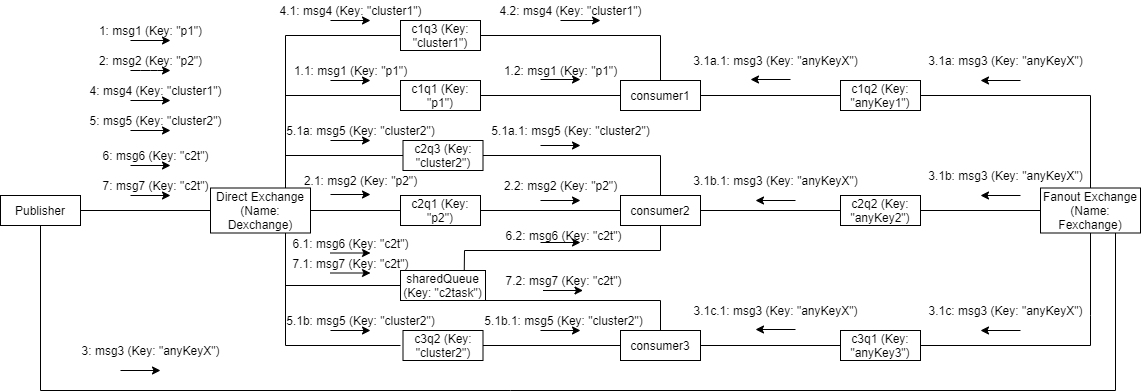}
\caption{\texttt{Messaging example}}%
\label{fig:RabbitMessageExample}
\end{figure}

\noindent
~\autoref{fig:RabbitMessageExample} has one publisher and three consumers. It deals with global and cluster broadcast, load-balanced tasks, and private messages (see the code in~\autoref{ce:RabbitMQMessage}, for running it install \textsc{Go}, \textsc{Erlang}, Git, and \textsc{RabbitMQ} server).

As shown in~\autoref{cod:MessagePublisher}, the publisher sends seven messages, six of them are sent to the direct exchange and one to the fanout exchange. 
There are six queues bound to the direct exchange (see~\autoref{cod:MessageConsumer1},~\autoref{cod:MessageConsumer2} and~\autoref{cod:MessageConsumer3}) with different bounding routing keys. One queue is bound to \texttt{cluster1}, the cluster broadcast queue of \texttt{consumer1}. 

Two queues are bound to \texttt{cluster2}: the cluster broadcast queue of the \texttt{consumer2} and the cluster broadcast queue of the  \texttt{consumer3}. Once the \textit{msg5} with the publishing routing key of \texttt{cluster2} has arrived at the direct exchange, the exchange sends it to the two queues concurrently.

The private queue of \texttt{consumer1} is bound with the key \texttt{"p1"}. The private queue of \texttt{consumer2} is bound with the key \texttt{"p2"}. The sharedQueue is bound with the key \texttt{"c2t"} to \texttt{consumer2} and \texttt{consumer3}, and all the messages of it are consumed in a load-balanced way.
There are three queues bound to the fanout exchange. The publishing and bounding routing keys are ignored, all the messages sent to the exchange are sent to all the queues bound to it concurrently.

The example in this section shows us how to communicate in a distributed system with private message, global broadcast, group broadcast, and group load-balanced task distribution using \textsc{RabbitMQ} queues, fanout and direct exchanges.
Another tool used in distributed message communication is \textsc{Mangos} that is next introduced.

\subsection{\textsc{Mangos}}

\textsc{Mangos} implements scalability protocols~\cite{url_nanomsg}, also called communication patterns. Currently it supports six patterns: req/rep, pub/sub, pair, bus, push/pull, and surveyor/respondent. It can handle the most common distributed operations, i.e.\ broadcasting and load-balanced distribution.
We present a push/pull example and a pub/sub one to show how \textsc{Mangos} operates. 

\begin{figure}[!htb]
\centering
\includegraphics[width=1\textwidth, height=3cm]{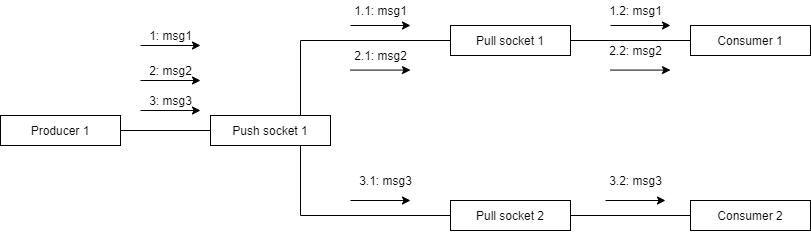}
\caption{Mangos push/pull}\label{fig:MangosPushPull}
\end{figure}

The push/pull pattern deals with load-balanced messaging. As~\autoref{fig:MangosPushPull} shows, the producer sends messages to consumers in a round-robin way (see code in~\autoref{ce:MangosPushPull}).

In~\autoref{cod:MGPipeConsumer1} and~\autoref{cod:MGPipeConsumer2} the pull sockets of the consumers separately connect to the push socket of the producer by the \texttt{Dial} method. They use the \texttt{Recv} method in a loop to receive the messages. In~\autoref{cod:MGPipeProducer}, the producer's push socket listens to a URL with the \texttt{Listen} method. The producer sends a message by the \texttt{Send} method.
The messages sent through the push socket are distributed to the connected pull sockets in a round-robin manner.

\begin{figure}[!htb]
\centering
\includegraphics[width=1\textwidth, height=5cm]{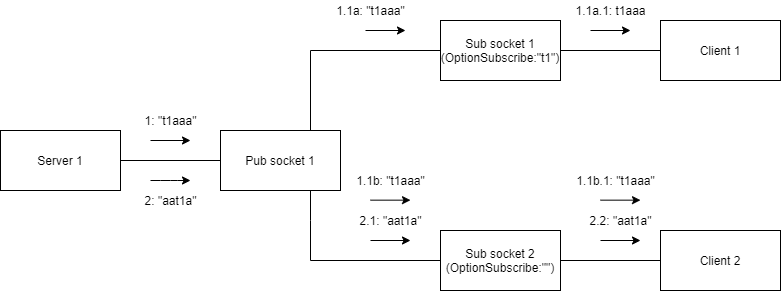}
\caption{Mangos pub/sub}\label{fig:MangosPubSub}
\end{figure}

The pub/sub pattern can handle the subscription and broadcasting. As depicted in~\autoref{fig:MangosPubSub} the server sends messages, while the two clients receive the ones that they are subscribed to (see code in~\autoref{ce:MangosPubSub}).

In~\autoref{cod:MGPubSubClient1} and~\autoref{cod:MGPubSubClient2} the sub sockets of the consumers separately connect to the pub socket of the producer by the \texttt{Dial} method. They use the \texttt{SetOption} method to set the subscription options before using the \texttt{Recv} method in a loop to receive the messages. The first client subscribes to the \texttt{"t1"}, which means it is interested in all messages that begin with \texttt{"t1"}. The second client subscribes to an empty string, which means it is interested in all the messages. In~\autoref{cod:MGPubSubServer} the pub socket of the producer listens to a URL with the \texttt{Listen} method. The producer sends the message by the \texttt{Send} method.
Once a message is sent through the pub socket, it is sent to all the sub sockets subscribed to it.

The example shows that any type of communication pattern can be chosen for implementing a distributed system. However, the \textsc{Mangos} patterns are separated, i.e.\ you need to import and create different sockets for each pattern separately. Therefore, too much code is needed when building a more complex topology of nodes.

\subsection{\textsc{RabbitMQ} and \textsc{Mangos} comparison}
Before introducing the features of the new project, \textsc{RabbitMQ} and the \textsc{Mangos} are compared from three different viewpoints. 

\textsc{RabbitMQ} is implemented in \textsc{Erlang} which has built-in distributed constructs. The mail system of \textsc{Erlang} follows the actor model, but it needs an extra runtime environment when executing the code.
\textsc{Mangos} is implemented in \textsc{Go} which has powerful built-in concurrency constructs like goroutines and channels. \textsc{Go} compiles the program into machine code and there is no need for extra runtime, like in case of a virtual machine or an interpreter.

\textsc{RabbitMQ} is broker-based. If its servers do not run in high-availability mode and the broker fails, then many of the nodes get affected and lose connection to the server. 
\textsc{Mangos} is brokerless, thus no broker can fail. However, without a broker, the producers and consumers know each other as they are directly connected.

\textsc{RabbitMQ} is user-friendly and it has built-in constructs (exchanges and queues) with different mechanisms. These constructs can be easily combined to meet the need of practical scenarios. The \textsc{RabbitMQ} direct exchange has subscription and broadcast features that are similar to the \textsc{Mangos} pub/sub pattern. The queues of \textsc{RabbitMQ} are load balanced, which is similar to the \textsc{Mangos} push/pull pattern. Broadcasting and load balancing can be easily combined in \textsc{RabbitMQ}, as shown in~\autoref{fig:RabbitMessageExample}. However, in \textsc{Mangos} separate patterns are used for each topology and they are not easy to mix. So the main drawback of \textsc{Mangos} is that the patterns are sharply separated. 

In practice, usually there are scenarios that need combined patterns. If you want to implement the example of~\autoref{fig:RabbitMessageExample} in \textsc{Mangos}, you need to create twelve sockets: three sets of push/pull sockets for the private messages of \texttt{client1}, \texttt{client2} and the task messages of cluster2; one set of push/pull sockets for the private messages of \texttt{client1}; three sets of pub/sub sockets for the global broadcast, the cluster1 and the cluster2 broadcast.

Both \textsc{RabbitMQ} and \textsc{Mangos} are powerful distributed communication tools. However, there is still room for improvements of usability and of scalability. In the next section, we introduce a new distributed communication model, that aims at increasing the efficiency of the above two features.

\section{New Sustainable Communication Model}

We are currently designing and implementing a new communication model and tool. This model takes advantage of some aspects of \textsc{RabbitMQ} and \textsc{Mangos}, and it also provides more features as enumerated here.

\textbf{Usability}. Our goal is to decrease the number of objects a programmer needs to create in the code of a communication scenario. The objects are different for each distributed communication tools: they are exchanges and queues of \textsc{RabbitMQ}; they are special sockets of \textsc{Mangos}; they are communicators of the new distributed communication project. The communication scenario to test different tools should cover private messages, global broadcast, group broadcast, and group load balance.
Compared to \textsc{Erlang}, there will be more routing information in the port mapping table. The new model sacrifices a bit of speed for more usability because the communicator needs to distinguish between the mechanisms to be applied. As a consequence, we can easily handle complex communication scenarios without the need of declaring many exchanges or creating many special sockets. All the messages are sent to the communicator with a routing information list that contains the routing mechanisms and routing tags.
    
\textbf{Scalability}. Our goal is to increase the number of connections the program can handle in a fixed time. Unlike the \textsc{Mangos}, the new model does not use separate ports for different mechanisms. One communicator uses one port and that is usually enough for each program as it can deal with different types of message-sending mechanisms.

\textbf{Strictly decentralized and symmetrical}. This model is designed to be socket-based, brokerless and it follows a modified actor model~\cite{actorModel}. The new communication tool is composed of multiple communicators that have equal roles. Each program of the distributed system uses a communicator to send and receive messages. In the actor model, the program and its communicator together form an \textit{actor}.

\textbf{Efficiency}. The goal is to decrease the time needed to transfer with various mechanisms fixed amount of messages. There is a trade-off between usability and efficiency in this model.  Although it needs more computations to handle different mechanism, efficiency should not be a weakness of this model. In other words, this communication tool may be slower than  \textsc{Mangos}, but not much slower.

\textbf{No extra runtime environment}. The new model design is implemented in \textsc{Go} which compiles the program to machine code. Thus, no need for extra runtime environment to run a program.
    
\textbf{Layered message packaging}. Only the necessary information is transferred. For example, after the routing operations, the complete routing information is not needed anymore.
    
\textbf{Automation}. Less work needs to be done by the programmer as there are dedicated goroutines handling routings, tag matchings, message pushing, and others.

The new model takes sustainability into account. It is lightweight; therefore, it is easy to maintain and upgrade. Due to its excellent usability and automation, less work needs to be done by the programmers and more energy can be saved.

\section{Conclusion}
In this paper we gave a brief overview of some  distributed communication tools and we elucidated the need of novel ones. For the Summer School 2022 more content will be included about the actor model and practical examples. We will also provide a more detailed comparison between different distributed communication models, as well as, we introduce more about the design and implementation details of our own new distributed communication model and tools. 

We will use slides and code examples as teaching materials at the lecture and the lab sessions of the summer school.

After studying the tutorial, the students will know how to handle distributed communication using \textsc{Mangos}, \textsc{RabbitMQ}, and a new sustainable distributed communication tool. They will acquire new skills and will be trained for practical usages of distributed communication tools.

\section*{Acknowledgements}
This  paper  acknowledges  the  support  of  the  Erasmus+ Key Action 2 (Strategic partnership for  higher education) project $No$ 2020-1-PT01-KA203-078646: “SusTrainable – Promoting Sustainability as a Fundamental Driver in Software Development Training and Education”. The information and views set out in this paper are those of the authors and do not necessarily reflect the official opinion of the European Union. Neither European Union institutions and bodies, nor any person acting on their behalf may be held responsible for the use which maybe made of the information contained therein.

\setcounter{section}{0}
\renewcommand{\thesection}{\Alph{section}} 
\section{Code listings and outputs}
Listings are not complete, they just illustrate the communication-related parts.

\subsection{RabbitMQ messaging example}\label{ce:RabbitMQMessage}

\begin{lstlisting}[language=Go,style=numbers,frame=single,label={cod:MessagePublisher},caption={Messaging example, Publisher}, captionpos=b]
//Publish Messages
publishMsg(ch, "Dexchange", "p1", "private msg1")
publishMsg(ch, "Dexchange", "p2", "private msg2")
publishMsg(ch, "Fexchange", "anyKeyX", "fanout broadcast")
publishMsg(ch, "Dexchange", "cluster1", "cluster1 broadcast")
publishMsg(ch, "Dexchange", "cluster2", "cluster2 broadcast")
publishMsg(ch, "Dexchange", "c2t", "cluster2 task1")
publishMsg(ch, "Dexchange", "c2t", "cluster2 task2")
\end{lstlisting}
\begin{lstlisting}[language=Go,style=numbers,frame=single,label={cod:MessageConsumer1},caption={Messaging example, Consumer1}, captionpos=b]
//Private messages
c1q1, err := ch.QueueDeclare("",false,true,true,false,nil)
err = ch.QueueBind(c1q1.Name,"p1","Dexchange",false,nil)
//Fanout broadcast
c1q2, err := ch.QueueDeclare("",false,true,true,false,nil)
err = ch.QueueBind(c1q2.Name,"anyKey1","Fexchange",false,nil)
//Cluster broadcast
c1q3, err := ch.QueueDeclare("",false,true,true,false,nil)
err = ch.QueueBind(c1q3.Name,"cluster1","Dexchange",false,nil)
\end{lstlisting}

\begin{lstlisting}[language=Go,style=numbers,frame=single,label={cod:MessageConsumer2},caption={Messaging example, Consumer2}, captionpos=b]
//Private messages
c2q1, err := ch.QueueDeclare("",false,true,true,false,nil)
err = ch.QueueBind(c2q1.Name,"p2","Dexchange",false,nil)
//Fanout broadcast
c2q2, err := ch.QueueDeclare("",false,true,true,false,nil)
err = ch.QueueBind(c2q2.Name,"anyKey2","Fexchange",false,nil)
//Cluster broadcast
c2q3, err := ch.QueueDeclare("",false,true,true,false,nil)
err = ch.QueueBind(c2q3.Name,"cluster2","Dexchange",false,nil)
//Task messages
sharedQueue, err :=ch.QueueDeclare("SharedQueue",false,true,false,false,nil)
err = ch.QueueBind(sharedQueue.Name,"c2t","Dexchange",false,nil)
\end{lstlisting}

\begin{lstlisting}[language=Go,style=numbers,frame=single,label={cod:MessageConsumer3},caption={Messaging example, Consumer3}, captionpos=b]
//Fanout broadcast
c3q1, err := ch.QueueDeclare("",false,true,true,false,nil)
err = ch.QueueBind(c3q1.Name,"anyKey3","Fexchange",false,nil)
//Cluster broadcast
c3q2, err := ch.QueueDeclare("",false,true,true,false,nil)
err = ch.QueueBind(c3q2.Name,"cluster2","Dexchange",false,nil)
//Task messages
sharedQueue,err := ch.QueueDeclare("SharedQueue",false,true,false,false,nil)
err = ch.QueueBind(sharedQueue.Name,"c2t","Dexchange",false,nil)
\end{lstlisting}

\begin{Go}
go run consumer1.go
Consumer1 waiting for msgs
Received: private msg1
Received: fanout broadcast
Received: cluster1 broadcast
\end{Go}

\begin{Go}
go run consumer2.go
Consumer2 waiting for msgs
Received: private msg2
Received: fanout broadcast
Received: cluster2 broadcast
Received: cluster2 task1
\end{Go}

\begin{Go}
go run consumer3.go
Consumer3 waiting for msgs
Received: fanout broadcast
Received: cluster2 broadcast
Received: cluster2 task2
\end{Go}

\begin{Go}
go run publisher.go
Sent:  private msg1
Sent:  private msg2
Sent:  fanout broadcast
Sent:  cluster1 broadcast
Sent:  cluster2 broadcast
Sent:  cluster2 task1
Sent:  cluster2 task2
\end{Go}

\subsection{\textsc{Mangos} push/pull example}\label{ce:MangosPushPull}
\begin{lstlisting}[language=Go,style=numbers,frame=single,label={cod:MGPipeProducer},caption={Mangos push/pull pattern example, Producer}, captionpos=b]
package main
import (
    "go.nanomsg.org/mangos/v3/protocol/push"
    _ "go.nanomsg.org/mangos/v3/transport/all"
)
func main() {
    var url = "tcp://127.0.0.1:40899"
    sock, err := push.NewSocket()
    defer sock.Close()
    err = sock.Listen(url)
    for _, msg := range links {
        err = sock.Send([]byte(msg))
    }
}
\end{lstlisting}

\begin{lstlisting}[language=Go,style=numbers,frame=single,label={cod:MGPipeConsumer1},caption={Mangos push/pull pattern example, Consumer1}, captionpos=b]
package main
import (
    "fmt"
    "log"
    "go.nanomsg.org/mangos/v3/protocol/pull"
    _ "go.nanomsg.org/mangos/v3/transport/all"
)
func main() {
    var url = "tcp://127.0.0.1:40899"
    sock, err := pull.NewSocket()
    err = sock.Dial(url)
    for { 
        msg, err := sock.Recv()
        if err != nil {
            failOnError(err,"cannot receive from mangos Socket")
        }
        fmt.Println("Consumer1 received: ", string(msg))
    } 
}
func failOnError(err error, msg string) {
    if err != nil {
        log.Fatalf("%s: %s", msg, err)
    }
}
\end{lstlisting}

\begin{lstlisting}[language=Go,style=numbers,frame=single,label={cod:MGPipeConsumer2},caption={Mangos push/pull pattern example, Consumer2}, captionpos=b]
    ...
        fmt.Println("Consumer2 received: ", string(msg))
    ...
\end{lstlisting}

\begin{Go}
go run producer.go
Producer sent:  http://google.com
Producer sent:  http://golang.org
Producer sent:  http://web0.com
Producer sent:  http://web1.com
Producer sent:  http://web2.com
Producer sent:  http://web3.com
Producer sent:  http://web4.com
Producer sent:  http://web5.com
Producer sent:  http://web6.com
Producer sent:  http://web7.com
Producer sent:  http://web8.com
Producer sent:  http://web9.com
\end{Go}

\begin{Go}
go run consumer1.go
Consumer1 received:  http://google.com
Consumer1 received:  http://web0.com
Consumer1 received:  http://web2.com
Consumer1 received:  http://web3.com
Consumer1 received:  http://web5.com
Consumer1 received:  http://web7.com
Consumer1 received:  http://web8.com
\end{Go}

\begin{Go}
go run consumer2.go
Consumer2 received:  http://golang.org
Consumer2 received:  http://web1.com
Consumer2 received:  http://web4.com
Consumer2 received:  http://web6.com
Consumer2 received:  http://web9.com
\end{Go}

\subsection{Mangos pub/sub example}\label{ce:MangosPubSub}
\begin{lstlisting}[language=Go,style=numbers,frame=single,label={cod:MGPubSubServer},caption={Mangos pub/sub pattern example, Server}, captionpos=b]
package main
import ( "go.nanomsg.org/mangos/v3/protocol/pub"
    _ "go.nanomsg.org/mangos/v3/transport/all" )
func main() {
    var url = "tcp://127.0.0.1:40899"
    sock, err := pub.NewSocket()
    defer sock.Close()
    err = sock.Listen(url)
    time.Sleep(10*time.Second)
    msg:= "t1aaaaa"
    _ = sock.Send([]byte(msg))
    fmt.Println("Server sent:", msg)
    msg = "t1bbbbb"
    _ = sock.Send([]byte(msg))
    fmt.Println("Server sent:", msg)
    msg = "bbbt1bb"
    _ = sock.Send([]byte(msg))
    fmt.Println("Server sent:", msg)
    msg = "t2ccccc"
    _ = sock.Send([]byte(msg))
    fmt.Println("Server sent:", msg)
    msg = "t2ddddd"
    _ = sock.Send([]byte(msg))
    fmt.Println("Server sent:", msg)
    msg = "t3ddddd"
    _ = sock.Send([]byte(msg))
    fmt.Println("Server sent:", msg)
}
...
\end{lstlisting}

\begin{lstlisting}[language=Go,style=numbers,frame=single,label={cod:MGPubSubClient1},caption={Mangos pub/sub pattern example, Client1}, captionpos=b]
package main
import (
    "go.nanomsg.org/mangos/v3"
    "go.nanomsg.org/mangos/v3/protocol/sub"
    _ "go.nanomsg.org/mangos/v3/transport/all"
)
func main() {
    var url = "tcp://127.0.0.1:40899"
    sock, err := sub.NewSocket()
    err = sock.Dial(url)
    err = sock.SetOption(mangos.OptionSubscribe, []byte("t1"))
    for {
        msg, err := sock.Recv()
        fmt.Println("Client1 received: ", string(msg))
    }
}
\end{lstlisting}

\begin{lstlisting}[language=Go,style=numbers,frame=single,label={cod:MGPubSubClient2},caption={Mangos pub/sub pattern example, Client2}, captionpos=b]
err = sock.SetOption(mangos.OptionSubscribe, []byte(""))
fmt.Println("Client2 received: ", string(msg))
\end{lstlisting}

\begin{Go}
go run server.go
Server sent: t1aaaaa
Server sent: t1bbbbb
Server sent: bbbt1bb
Server sent: t2ccccc
Server sent: t2ddddd
Server sent: t3ddddd
\end{Go}

\begin{Go}
go run client1.go
Client1 received:  t1aaaaa
Client1 received:  t1bbbbb
\end{Go}

\begin{Go}
go run client2.go
Client2 received:  t1aaaaa
Client2 received:  t1bbbbb
Client2 received:  bbbt1bb
Client2 received:  t2ccccc
Client2 received:  t2ddddd
Client2 received:  t3ddddd
\end{Go}
\renewcommand{\thesection}{\arabic{section}} 

\title{Teaching Green Computing for the Internet of Things}
\author{%
    Mart Lubbers\orcidID{0000-0002-4015-4878} \and 
    Pieter Koopman\orcidID{0000-0002-3688-0957} 
}
\authorrunning{M.\ Lubbers and P.\ Koopman}
\institute{Institute for Computing and Information Sciences, Radboud University Nijmegen, The Netherlands
\email{\{mart,pieter\}@cs.ru.nl}}
\date{Sustrainable Teacher Training: November 1--5, 2021}
\maketitle
\begin{abstract}
The Internet of Things (IoT) is booming and is largely powered by microcontrollers.
While microcontrollers use little energy, the sheer amount of them consume quite a lot, and as they often operate on batteries they produce a lot of e-waste.
Task-oriented programming (TOP) is a novel declarative programming paradigm that offers apt abstractions for workflow and is suitable for the IoT as well.
The mTask system is a TOP framework designed specifically for the edge layer of IoT applications and offers abstractions for things such as peripheral access and data serialisation.
Moreover, due to the execution semantics of tasks in the system, a clever scheduler can run multiple tasks in an interleaved fashion and utilise the low-power sleep modes native to microprocessors to the fullest.
mTask is fully integrated in iTask, a TOP system for multi-user distributed web-based workflow applications, making it possible to program all layers of an IoT system from one declarative specification.

In the lecture  the students will incrementally learn about and experiment with reducing the power consumption on microcontrollers.
During the lab session, they will use the mTask system on actual microcontrollers as a rapid prototyping tool.
In an incremental fashion, the power consumption of a typical IoT application will be reduced.
During the process, they verify their results using a current measurement sensor that they access via its web interface.

We can assume limited knowledge upfront. 
This form of green computing requires knowledge of topics like IoT, microcontrollers and TOP.
Hence, we have to make firm decisions about what we can tell and to which amount of detail.
\end{abstract}

\begin{keywords}
teaching \and
internet of things \and
embedded systems \and
green computing \and
task-oriented programming
\end{keywords}


\section{Internet of Things}
At the time of writing over 12.3 billion edge devices are powering the Internet of Things (IoT)\footnote{IoT Analytics\\ \url{https://iot-analytics.com/product/state-of-iot-summer-2021/}, accessed on 2022-02-14.}.
Often, IoT systems are designed using layered architectures.
The typical four-tier architecture consists of a presentation layer; an application layer; a network layer; and a perception, or edge, layer~\cite{muccini-iot-architectural}.
The presentation layer provides the human system interaction using apps, websites or other modalities.
The application layer contains the core of the system, the data aggregation, processing and storage.
This layer is typically powered by high level applications running on powerful servers.
At the edge of the architecture resides the perception layer, powered by microcontrollers equipped with sensors and actuators.
Connecting all this is done by the network layer.
It provides the firmament between the application and perception layer using ordinary protocols such as GSM or WiFi but also specialised IoT technologies such as Bluetooth Low Energy, Zigbee or LoRa.
Every layer is programmed using different abstraction levels, paradigms, and programming languages and operates on different platforms and hardware architectures resulting in semantic friction.

It is vital to reduce the power consumption of the microprocessors powering the edge or the perception layer~\cite{popli_green_2021}.
While the microprocessors powering the edge layer of an IoT system do not consume a lot of energy, the sheer number of them totals to quite the amount.
Furthermore, many edge devices operate on batteries eventually resulting into e-waste when they run out since recycling or even replacing them is often not possible.
Microprocessors typically use a fairly constant amount of power under normal circumstances.
Nevertheless, most microprocessors boast specialised low-power sleep modes to reduces the power consumption significantly.
The key to reducing the energy consumption is therefore to sleep as long and as deep as possible~\cite{6577767}.
To reach these low-power requirements, the microprocessor disables many parts of the chip to save power.
For example, pins are lose power; peripherals, such as the analogue to digital converter (ADC), are turned off; the random access memory (RAM) is left without power, thus wiping the contents; and communication may be suspended or stopped.
External events, such as hardware interrupts from pin I/O or a real-time clock are ordinarily required to wake the device up again.
Maximising the sleeping time can be achieved by clever scheduling, reducing the polling frequency of sensors and replacing polling my hardware interrupts.

\section{Task-oriented programming}
Task-oriented programming (TOP) is a recently emerged declarative programming paradigm.
In TOP, tasks represent work and are the basic building blocks that can be combined using combinators to describe workflow~\cite{plasmeijer_task-oriented_2012}.
The declarative specification of a program only describes the \emph{what} and not the \emph{how} of execution.
The system that executes the tasks makes sure that details are taken care of all abstractions that were made such as the user interface, data storage and communication~\cite{wang_maintaining_2018}.
Tasks are often modelled as event driven rewrite systems that have an observable value that can be used to communicate between related tasks.
Moreover, data can be shared between unrelated tasks using shared data sources (SDSs).

There are several TOP languages and systems for executing programs.
The oldest implementation is iTask, a domain specific language (DSL) embedded in the purely functional programming language Clean~\cite{brus_clean_1987} that grew from escapades made in generic programming~\cite{plasmeijer_itasks:_2007}.
From this TOP specification, iTask generates a multi-user, multi-platform, distributed web application to coordinate and execute the work.
There are also TOP languages with other purposes than executing the work, an example of this is TopHat, a formal foundation for TOP embedded in an enriched lambda calculus~\cite{steenvoorden_tophat_2019}.

mTask is a TOP language designed for the edge layer of the IoT that can, in conjunction with iTask, be used to program all layers of the IoT from a single source.
Where iTask abstracts away from gritty details encountered in web applications, mTask abstracts away from microprocessor details such as peripheral access, serialisation, communication, and multi-threading~\cite{lubbers_multitasking_2019}.
These abstractions result in less semantic friction and thus lower development and maintenance effort~\cite{lubbers_tiered_2020}.
The mTask language is implemented as a tagless-final style~\cite{carette_finally_2009} embedded DSL in Clean and iTask.
Using only few library tasks, devices can be connected in iTask, mTask tasks can be lifted to operate as iTask tasks, and SDSs from the server can be shared with the devices~\cite{lubbers_task_2018}.
When an mTask task is lifted, the TOP specification is compiled at run time to specialised byte code that is in turn sent to the device for interpretation~\cite{lubbers_interpreting_2019}.
The RTS on the device receives the byte code and executes the tasks accordingly, making sure that updates to SDSs and task values are communicated back to the server.
Since all the work is done at run time, the microprocessors running the mTask RTS only need to be programmed once, after which they are able to constantly receive new programs tailor-made for the work that needs to be done.

\section{Course information}
Recently an abstraction for many power consumption influencing features such as sleep mode utilisation and hardware interrupts were added to the mTask system~\cite{crooijmans_reducing_2021}. 
For the summer school we plan to give a lecture including a lab course on green computing for the IoT where we use these features to reduce the power consumption of a typical IoT system.
The gist of the course is to give students insight in state-of-the-art programming languages and paradigms while getting hands on experience with microprocessors without the hassle of setting up a tool chain, flashing chips and tinkering with wires and components.

\subsection{Lecture}
The lecture will start with the necessary background material for the lab session by answering the following questions:
\begin{itemize}
    \item What is the IoT?
    \item Why is it required to reduce the power consumption in IoT edge devices?
    \item How can we save power in IoT devices?
    \item What is TOP, the mTask system and the iTask system?
    \item What green computing functionality is there in mTask?
\end{itemize}

\subsection{Lab session}
In the lab session, the students are provided with two ready made, assembled, and programmed microprocessors to experiment with.
The first device is a microprocessor equipped with a current consumption sensor that constantly communicates the current consumption to the user via a web interface.
Using this device, the students can verify their effort for achieving power consumption reduction in their project.
The second device is a microprocessor equipped with some sensors and loaded with the mTask RTS that automatically connects to the local WiFi network.
On their own laptops, the students download and set up an mTask distribution that they can use to program the microprocessor using the mTask DSL.
The program that they receive is a typical IoT program that interacts with sensors, does some sensor fusion and is operable using buttons.
The students will then, in an incremental fashion, apply changes to the provided program to reduce the power consumption while maintaining functionality.
For example, they might change a task that polls a button to a task that uses interrupts to get notified of a state change or change the polling rate of the sensors to optimise the time that the microprocessor sleeps.
During these step by step tweaks to the program, the students constantly monitor the power consumption of the microprocessor using the other microcontroller for verification.

\section*{Acknowledgments}
This paper acknowledges the support of the Erasmus+ Key Action 2 (Strategic partnership for higher education) project No. 2020-1-PT01-KA203-078646: ``SusTrainable - Promoting Sustainability as a Fundamental Driver in Software Development Training and Education''.
The information and views set out in this paper are those of the author(s) and do not necessarily reflect the official opinion of the European Union. Neither the European Union institutions and bodies nor any person acting on their behalf may be held responsible for the use which may be made of the information contained therein.
Furthermore, this research is partly funded by the Royal Netherlands Navy.

\title{Algorithms for Sustainable System Topologies}
\author{Tihana Galinac Grbac\orcidID{0000-0002-4351-4082} \and
Neven Grbac} 
\authorrunning{T. Galinac Grbac and N. Grbac}
\institute{Juraj Dobrila Univeristy of Pula, 
Zagreba\v{c}ka 30, HR-52100 Pula, Croatia \\ 
\email{tihana.galinac@unipu.hr, neven.grbac@unipu.hr}
}
\maketitle
\begin{abstract}
Sustainability has become one of the global goals for further technology evolution.
Technology evolution is implying automation in every aspect of human operation with help of software. Smart software solutions can help reach the aforementioned global goal. However, smart modeling and design solutions have to find the balance between sustainable environmental goals and software quality goals. Current software and system engineering practices, especially within the system modeling and design phase, lack the engineering of this particular environmental aspect of software systems. When topological data analysis (TDA) is applied to the software system structures, we can observe numerous related quality attributes. In this lecture, we will provide details on how the topological data analysis can be used to build sustainable software structures that are in balance with the software quality model, discuss related implementation details, and introduce students to open source tools, in which they can experiment with these algorithms.

\end{abstract}
\section{Introduction}%
\label{sect:Intro}
In all EU strategies there is clear orientation towards creation and adaptation to sustainable infrastructures and ecosystems in various sectors (manufacturing, tourism, healthcare, agriculture) through adoption of new technologies. Dynamic and unpredictable environmental conditions need fast and adequate product alignment in order to fulfill economic, social, technical and environmental sustainability. The ICT sector is clearly contributing to these goals and is applicable in almost every industry while sustainability is defined as `capacity to endure and preserve the function of a system over an extended period of time'~\cite{ICT4S1,ICT4S2}. Software, as the key ICT ingredient, drives the next technology evolution that would bring new opportunities for those various economy sectors such as tourism, manufacturing, government, health, education, etc. Therefore, as research and educational professionals, we focused our efforts towards developing and promoting sustainable goals within software engineering projects and innovations at the Juraj Dobrila University of Pula. Our way of action is critical for local society since our role in educating, fostering and developing open innovation platforms has significant impact in engaging whole local society involving local industry, government and public sector and enabling them efficient and effective alignment towards sustainable goal fulfilment.  

Number of researchers have reported that there is no clear understanding of the relation between sustainability and software quality attributes. Balancing among software quality attributes has  motivated numerous research studies so far, but still there are no clear and definite conclusions that can be precisely defined and uniformly stated for every software development environment~\cite{LagoGreen}. The impact of achieving software quality goals on sustainability has been weakly explored and future research efforts with numerous empirical studies have to explore and better understand these relations. This knowledge would be crucial in developing new technologies that would provide sustainable solutions to its users and define solid sustainable development models based on reliable sustainable indicators~\cite{FramingSustainability}. In order to contribute to this aim, we foresee that the new generation of software professionals should be timely educated to increase their awareness how their profession may contribute to fulfill the global sustainability goals. Therefore, we take our efforts to prepare course materials that can be incorporated into traditional software engineering curricula for educating new generation of software engineering professionals, which would allow them to better asses their software engineering decisions and their impact on global sustainability goals.

\section{Sustainable Software Topologies}

\subsection{Context of the lecture}

As concluded from extensive survey~\cite{LagoSurvey}, one of the most important quality attributes that impacts directly sustainability is modifiability, along with functional correctness, availability, interoperability and recoverability, which contributes to endurability of software systems. Therefore, in modern software engineering industries, continuous integration and continuous delivery development (CI/CD) practice is becoming one of the widely recognised and accepted norms of software system development. Such established development and integration environment, in which software modules are continuously developed, integrated into working software structures and delivered on Cloud infrastructure, open number of possibilities for achieving sustainable goals for software users, as well as software developers. Clearly defined software structures are prerequisites for such practice and it is still the major concern within software development community. Insufficient understanding, lack of knowledge and awareness of all possible solutions and its threats is the major obstacle of CI/CD practice. \emph{Software structuring} problem would imply analysis of local software attributes in relation to global software attributes and their ability to endure within the environmental conditions. This problem is an upgrade of the software architecting problem, but in wider context which has opened with innovations in ICT and network technologies. 

From the traditional software engineering perspective, the software architecture is the key artefact used in visualising the system, creating, assessing and communicating decisions about its modifications and evolution during the whole software lifecycle. Therefore, our emphasis here is on software architecture as the main software engineering artefact for managing software system and it refers to the fundamental structures of a software system and the discipline of creating such structures and systems.

We observe here complex software system as a modular system composed of components with several hierarchical levels of abstraction. We distinguish among global system properties, which describe system behaviour that we can measure on a running system and may be directly related to sustainable properties (such as reliability, availability, safety), and local component properties, which are measured on each system component statically or on running system and could be used to describe component behaviour (such as lines of code, cyclomatic complexity, cognitive complexity, number of interactions, number of defects). One of the key properties of such complex systems is that it is impossible to derive simple rules from local properties towards global properties. This complexity is the main reason why there exist numerous research efforts trying to understand relations among software quality attributes. 

In this lecture we represent complex system structure as a graph. Any (non-trivial) software system consists of components. Depending on the context and the purpose of the study, the components may be modules, subroutines, functions, units, classes, objects, \ldots The components may be viewed as vertices of a graph. The connections between components may be viewed as edges of a graph. These together form a graph representing a software system.

\subsection{Goals we study}

Our goal in this lecture is to study complex software systems and system topology with emphasis on sustainability and its relation to achieved system quality (modifiability, reliability and safety) while system is in operation. The source of our knowledge base for producing these materials is taken from the research project funded by the Croatian Science Foundation entitled ``Reliable and Safe Complex Software Systems: From Empirical Principles to Theoretical Models in View of Industrial Applications''. Within this project we study the research question ``How can the software structure analysis improve software engineering theory and practice having sustainable goals in mind?''. More particularly, the key problems we analyze are:
\begin{itemize}
    \item Can we develop foundations on sustainable software behavior?
    \item How can we measure software behaviour in the network?
    \item Can we predict and simulate software behaviour in the network?
    \item How to manage complex software system and its evolution?
    \item Are we able just by observing properties of system parts to predict and model its overall behaviour?
\end{itemize}
The way of action here is to develop models and tools that would enable us to better understand software structure and its dynamics and their influence on sustainability, reliability and system safety properties. Furthermore, we investigate how we could use software structure as a tool for modeling system behaviour. This is motivated by our earlier studies, which were based on the idea to represent software systems as graphs~\cite{SoftwareGraph}. There we derived some interesting conclusions about structural topology evolution of software systems and its relations to system size maturity and defectiveness~\cite{Petric}.

\subsection{Topological Data Analysis algorithms}
One of the very well established and applied mathematical theory is Topological Data Analysis (TDA) that we apply on software structuring problem. There are already numerous applications of TDA qualitative methods of topology to problems of machine learning, data mining and computer vision that were reported some successful results in tumor detection or spread of influenza. Topology identifies a global structure by reconstructing global structure from local substructures and enables studying relation of local and global system properties. Sometimes in the study of real world behavior it is not important to have precise quantitative models. In such cases the interest is in the shape, trends etc., and not in size, scale, time-frame~\cite{TDA}. In the sense of software systems that should be developed for sustainability whereas modifability is one of the key recognised quality attributes, this particular technique may provide us with better understanding about persistence of software quality attributes along with these dimensions such as size, scale and time-frame. Thus, the topological models of reality are qualitative, not quantitative. 

Graph of a software system may be viewed as a skeleton of a more general object – a topological space. This point of view allows the study of complex software systems topological structure and their behavior use of various topological methods and algorithms such as topological data analysis, algebraic topology approximate the structure and behavior of a complex system by a family of (less complex) topological objects. It is something like using (less complicated) derivatives and Taylor polynomials to model complicated functions in mathematical analysis.

From such an approach we may derive numerous benefits in terms of sustainable wise solutions. Let us mention few: structural software analysis (static and dynamic), resilience to failures (energy shocks), impact analysis, software redesign, dangerous structures (antipatterns), software classification, security mechanisms (energy threats) and many many others.

\section{Software Engineering Courses and Projects at the Juraj Dobrila University of Pula}
Within the undergraduate and graduate university study of computing at the Faculty of Engineering and the graduate study of informatics at the Faculty of Informatics at the Juraj Dobrila University of Pula, we engage student projects in collaboration with local government and key industrial partners on practicing various topology algorithms on software structures for understanding fulfilment of sustainability goals in various application domains. 

\subsection{Software engineering course and projects at the undergraduate and graduate study of computing}

Software engineering course is the capstone course at the undergraduate level of computing. This course aims to unify the knowledge acquired through the prerequisite courses of Operating Systems, Computer networks, Programming, Algorithms and Data Structures, and Object Oriented Programming. The course is organized in 30 hours of lectures, in which students follow oral lectures on the topic of software engineering based on \cite{Vliet}. These are introduction into software engineering discipline, software complexity, software management, software lifecycle models, software lifecycle phases from the requirements, design, implementation, testing, integration and verification. Throughout the lectures the emphasis is on collaborative models, methods and tools used to engineer complex software systems, open standards and innovation ecosystems. Furthermore, students have 30 hours of lectures from industry related to clean code solutions exercising with object oriented programming, experimenting with various design tools, get acquainted with the cloud platform following exercises presented at the previous Summer schools~\cite{GalinacRole1,GalinacRole2} and using locally configured Cloud based on Open Stack, and finally they are introduced to the continuous integration and continuous delivery paradigm, and open source repositories (GITHub). 

Based on the knowledge acquired during whole undergraduate study and these few final exercises, the students are ready to apply and experiment with TDA algorithms for sustainable software solutions. During the project, the students usually need knowledge acquired through Operating Systems and Computer Networks courses to understand how to configure and implement their algorithms within Cloud environment, where they need to understand how to configure IP addresses of their virtual machines on which the projects are running. Furthermore, they need programming, data structure and algorithms knowledge to develop efficient and sound algorithms and programs. Finally, since the algorithms are developed for software design, the students usually base their solutions on open source solutions available in GIThub. The major obstacle here is usually devoted to finding adequate tools and running them within the Openstack environment.   



\subsection{Complex systems engineering course and projects at the graduate study of informatics}

This course is an elective course at the final year of the graduate study of Informatics and Business Informatics. Majority of students already work or at least prepare to start working in some company, while finalising their graduate study. Our aim in this course is to provide them as much as possible with knowledge that is ready to apply within industrial context and that could support local industry to achieve operational excellence and sustainable goals. 

Therefore, we start engaging students into complex systems design applying the lectures provided on previous Summer schools~\cite{GalinacRole1,GalinacRole2}. From these lectures students are engaged to think about complex systems and during the lecture they are exercised to analyse the software system on which they currently work as a complex system. Furthermore, they are encouraged to recognise how to apply design principles of layering, hierarchy, modularity and abstraction. They are introduced to Cloud technology, continuous integration and continuous delivery solutions that can help in automatising part of software development lifecycle and thus reduce operational costs and provide sustainable development environment. During the course students work on their projects, in which they work on application of topology structure algorithms to achieve sustainable goals in variety of application domains such as tourism, e-commerce, business processes, e-governments and smart cities. 

\section{Content of the summer school lecture and educational goals}

The summer school lecture will be divided into three logical parts. The first part will provide the introduction to the sustainable software engineering, the second part will provide the introduction to the topological data analysis (TDA). Finally, the third part will provide exercises for the students, in which the students could practice the application of the TDA algorithms for sustainable software engineering.

The introduction to the sustainable software engineering will start with the definition of the key terms needed to follow the lecture. Firstly, we will define the sustainability from the software engineering perspective as defined in Sect.~\ref{sect:Intro}. Then, we will revise the software lifecycle from the perspective of artefacts developed in each phase, processes, methods, tools and roles involved. We will put special attention on the software structure as the main artefact used across all lifecycle phases. The effects of these will be examined from the perspective of the previously defined sustainability attributes, and these will be used to develop the key performance indicators (KPI) we will be using while practicing with the TDA algorithms.

The TDA part of the lecture will start with the formal definition of a graph, and in particular, how to obtain the graph from the software code at different levels. The software graph, as any graph, may be viewed as a simplicial complex over the two-element field $\mathbb{Z}_2$. The simplicial complex is a skeleton of a topological space, so that the software can be studied from the point of view of algbraic topology. In the lecture, we will briefly recall the definition of the simplicial complex and its linear algebra representation. We will explain some basic linear algebra over $\mathbb{Z}_2$ required to calculate basic topological objects associated to the software simplicial complex. 

Finally, in the last part, we will provide practical details regarding the application of the TDA theory on practical computational assignments evaluating sustainable key performance indicators on software graphs. As the programming language, we will use Python and explain how the graph structure may be imported and defined as a Python object. The main point here is that the students understand how software graphs may be imported into Python. The next assignment would be to practice with simple graph algorithms that may be used to describe software graphs. Finally, the student will practice with the TDA Python algorithms. Here we will ask them to reflect the obtained results in relation to software graph description, but also the experiences from the software engineering practice. We will discuss how TDA algorithms may be used to improve decision processes and automate some processes within software lifecycle.   

For the student assignment we assume that students have computer (e.g.\ bring their own laptop) and provided Internet connection, preferably Eduroam. Part of the exercises the students may perform on their local computers. However, due to problems that may arise when installing required packages and the amount of resources needed for TDA computation, we decided to provide ready to use virtual machines on the Cloud environment locally hosted at the University of Pula.

The main educational goals are the following:
\begin{itemize}
    \item definition of sustainable KPI within software lifecycle,
    \item use software structures for measuring KPI success,
    \item learn how to apply TDA on software structures,
    \item learn Python packages and algorithms for TDA. 
\end{itemize}

\section{Conclusion}

Software architecture is the main artefact used to communicate among various actors involved in software lifecycle phases and actors interested in software use. Complexity of modern software systems impose lack of understanding how to build software architectures that would produce software system behaviour acceptable for its users, that would allow timely adaptation to its environmental conditions and to fulfill the goals of its sustainable future evolution. We may overcome the huge uncertainty imposed by this lack of knowledge only with progressive research, education and solid models for empirical studies. In this lecture, we offer to the students tools to experiment with various software structures, apply various topology data analysis models which would enable them to better understand the relations of local and global software properties, relations of software structures and software quality attributes. These models are directly connecting software quality attributes to software sustainability goals. Along with this exercise, the students may incorporate within this modelling approach the energy depth measure and green software measures (presented by other lectures within this summer school) as local software attributes and relate them to other software quality attributes and sustainability goals with help of topology data algorithms offered within this lecture. The experimentation with tools would be secured over the Cloud environment hosted by the Juraj Dobrila University of Pula that the students may access over the Internet.

\section*{Acknowledgements}

This paper acknowledges the support of the Erasmus+ Key Action 2 (Strategic partnership for higher education) project No. 2020–1–PT01–KA203–078646: “SusTrainable - Promoting Sustainability as a Fundamental Driver in Software Development Training and Education” and the support of the Croatian Science Foundation under the project HRZZ-IP-2019-04-4216. The information and views set out in this paper are those of the author(s) and do not necessarily reflect the official opinion of the European Union. Neither the European Union institutions and bodies nor any person acting on their behalf may be held responsible for the use which may be made of the information contained therein.

\title{Teaching by Example}
\titlerunning{Temporal-Logic Automated Solutions for Sustainable Educational Progress}
\author{Vladimir Valkanov\and Mihail Petrov}
\authorrunning{V. Valkanov, M. Petrov}
\institute{University Of Plovdiv ``Paisii Hilendarski'', \\ Faculty of Mathematics and Informatics, \\
24 Bulgaria Blvd., 4027 Plovdiv, Bulgaria \\
\email{vvalkanov@uni-plovdiv.bg}\\
\email{mihailpetrov@uni-plovdiv.bg}}
\maketitle
\begin{abstract}
The process of teaching and learning new knowledge can rarely be classified as standard. Teaching curriculum and the initial professor assessment of the audience are often indirectly related to the end result. The achievements which may vary based on discriminative factors such as gender, age, social background, past experience, learning through techniques such as listening, reading et al can be used as the basis of an analysis with the goal of identifying the factors allowing the students to achieve better results in the learning process~\cite{r1}.
\end{abstract}

\section{Introduction}

If a judgment is to be made as of today regarding one of the directions mankind has taken, without a doubt it has to do with civilizing our everyday lives. Each aspect of our everyday activity is invariably dependent on active interaction with electronic systems which make our activities easier. 

Against the backdrop of this technological scope, people spend a large part of their lives in an analogue medium among their peers in search of new knowledge. Education is one of those important ideas in our everyday lives that we dedicate large portions of our lives~\cite{r2} but only in the last few years it has become possible to notice a sufficient change in the manner in which it is absorbed each and every day. 

Without a doubt, today the classroom model is still almost unchanged from the days of the early schools; the teaching techniques and the subject matter have remained the same over time. But in the digital-culture era modern means of consuming such matter are making their way which can be mainly linked to the presence of the internet in our everyday lives. 

Dynamic changes in technology and the ever-more globalized reality of modern society provide for the increasing popularity of platforms for sharing of knowledge and skills in a formal, professional and academic manner. Examples for such solutions include:
\begin{itemize}
	\item  Khan Academy~\cite{r14};
	\item  Udacity~\cite{r12};
	\item  Udemy~\cite{r13};
	\item  Coursera~\cite{r11}.
\end{itemize}

The common thing among those platforms is how they combine high teaching potential with the capabilities of the modern internet to present short and meaningful educational materials in the form of text or video. Such methods are making way for formal education which gives us insight into inter-discipline knowledge and skills both from a scientific and from a professional point of view~\cite{r3}. 

\textbf{The common thing among the above systems is their idea of adaptiveness} based on the information and the time users spend on them. Digitalization does not just change the media we use every day but also creates connections among multiple types of media. The individual style of learning and understanding information, our individual educational needs are all aspects that modern digital platforms focus on.

\section{Application of systems based on temporal models}

The concept of temporal sequences can be easily found in different stages of the development of modern technologies. 

Temporal analysis is mainly used in the management of investment decisions based on market cycles shaped by the changes in the values of currency pairs, precious metals, raw materials, essential goods as well as obligation tools and indices. Each change in the market priority can be modelled as a change in the decisions made in a situation~\cite{r4} requiring positive analysis of a given situation or one which is aimed towards risk reduction. The keyword which we will use in this context is adapting a strategy based on temporal shift.

Analyzing temporal models in this particular meaning can be found in the management of risk situations having to do with the maintenance of information systems which has direct relation to managing critical situations. Managers dealing with those types of systems need to constantly analyze the information stream in order to find an adequate solution in optimizing their strategy.

In the context of the given examples, the modern educational system can also be counted towards the adaptive systems based on temporal sequences. Modeling knowledge in the context of an open system is a function of several variables, including:
\begin{itemize}
	\item  Initial knowledge that the student has;
	\item  Set of specific characteristics which can give them competitive advantage in acquiring specific skills -- including physical predisposition; cultural, linguistic or mental characteristics;
	\item  Time spent in adapting the information stream into knowledge -- through active application of a set of educational approaches.
\end{itemize}

The main focus of our research is directed towards reviewing temporal sequences in describing mechanisms for adapting educational processes to the needs of a large set of educational agents. 

\section{Active and passive interaction with the educational platforms}

The way in which a user communicates with the platform interface gives the most important information necessary for building a behavior model for the user which is the most important source of information for the system's personalization and adaptation. Depending on whether the user is actively performing certain physical actions or is in a state of rest or delay could give different information to the system for the current state of mind of the agent. 

In order to be able to determine the user's actions we first need to get to know the instruments available in the platform for user interaction. 
\begin{figure}
\begin{center}
\includegraphics[width=0.6\textwidth]{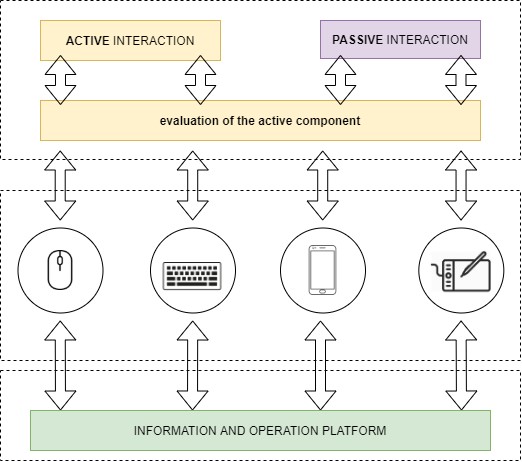}
\caption{Intercepting and assessing the interaction component through interpretational instruments.}\label{fig1}
\end{center}
\end{figure}

\textbf{Interaction instruments with the platform interface:}

\begin{itemize}
	\item  Mouse and keyboard;
	\item  Mobile-device capacitive display;
	\item  Graphic pad for generating graphic primitives through an electronic stylus;
	\item  Peripherals generating audio waves;
	\item  Enhanced-reality peripherals;
	\item  Terminal interfaces for managing current user states.
\end{itemize}

\textbf{Active interaction}

The most popular type of interaction has to do with the agent's active participation in the processes described by the system. Active interaction can be classified depending on the events which it produces:
\begin{itemize}
	\item  Verification interaction;
	\item  Aggregation interaction. This is a process of generating events based on extracting information;
	\item  Interaction directed towards acquiring new material.
\end{itemize}

Generated events serve as the basis of building an information-operative platform which realizes the physical connection between the system and the real actions of the user.

\textbf{Passive interaction}

Passive is any interaction where the user does not deliberately generate results from their own actions or inactivity. In the context of the researched subject several passive activities are measured which result in the aggregation of additional information. 

It is important to point out that passive interaction cannot be measured with absolute preciseness and it is necessary to approach the acquired data with a certain dose of skepticism.

Reading and reflecting are passive events where the user reads and reflects information in the form of text, meaning lessons or problems that need to be solved. In order to account for such a passive action, the following steps need to be present:
\begin{itemize}
	\item  The user has loaded a particular lesson or problem;
	\item  The user spends time analyzing the contents;
	\item  The user is navigating through the problem or lesson in order to get acquainted to additional details having to do with the analyzed material. 
\end{itemize}

\section{Categorizing the incorrect state severity}

When we make mistakes, we learn to better understand the world around us. In the programming environment, making mistakes helps us better recognize the tools we work with, namely programming language, code technology and software application architecture. We can use the sequential classification in order to describe the system~\cite{r7}.

This is a classification based on the sequence of the incorrect and correct states of the problem. The solution of the problem can be divided over several simple steps necessary for solving a more complex task that comprises the whole solution. In other words, the solution generally comes down to following an algorithm to find the proper metrics and reach the answer. Math problems are often an adequate example of problems that can be classified sequentially:
\begin{itemize}
	\item  The number of steps necessary for finding the result is often determined by the type of task;
	\item  To progress further, you must break the problem into small micro-tasks generally required by the previously solved tasks;
	\item  The result depends on the solutions of the previous tasks.
\end{itemize}

The essential part of the sequential classification is the labeling of every single incorrect state to classify the frequency and the severity of every single one of them. We are going to use one example of an incorrect state that is defined in this paper. Over the process of solving a task, the user can make the following mistakes:
\begin{itemize}
	\item  Syntax;
	\item  Requirement;
	\item  Semantical;
	\item  Platform-specific;
	\item  Operational;
	\item  Run-time.
\end{itemize}

Every single one of these mistakes is defined as an incorrect state and it is a part of a sequential graph labeled with the type of the invalid state and the state of the task. In the sequential classification, we are interested only in the chain of invalid states and not the time between the states. The process of running over multiple invalid states of the same type is labeled as a syntactical cluster and is analyzed according to its state~\cite{r5}. There may be many reasons why agents encounter a cluster of invalid states, for example:
\begin{itemize}
	\item  Not understanding the syntax of the language;
	\item  Not understanding the flow of the program;
	\item  Not understanding the language constructs.
\end{itemize}

Naturally, not all syntax errors can be assigned to the same category, the following groups can be defined:
\begin{enumerate}
	\item  An inadvertent syntax error -- they are usually associated with the omission of a letter when writing a program construct, the omission of a space character, or a termination symbol. The concept that can be used to analyze such a situation is called Levenshtein distance for measuring the distance between the keystrokes of the keyboard and classifying the mistake that the agent made~\cite{r9}.
	\item  Syntax errors made by a misunderstanding of language constructs -- if the agent has previous experience with another programming language, the connection between their previous experience and the current code. The use of language constructs atypical of the language under study can be readily recognized using language compilers. Such errors demonstrate the need for additional guidance on tasks that support the use of language constructs typical of the current programming language~\cite{r10}.
	\item  Syntax errors not classified in the above two categories. If a linguistic error cannot be determined, then it is result of a much more complex error in the understanding of matter. Getting an agent into a cluster of invalid states can immediately indicate a global problem in understanding of the matter or language constructs. The sequence of incorrect states produces a graph (Fig.~\ref{fig2}).
\end{enumerate}

\section{Classification based on time intervals}

\begin{figure}
\begin{center}
\includegraphics[width=1\textwidth]{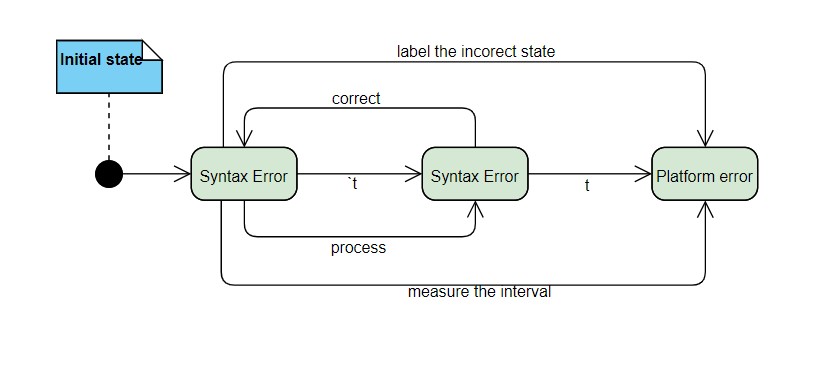}
\caption{Linking the incorrect states as an ordered list by categorizing their severity.}\label{fig2}
\end{center}
\end{figure}

In figure~\ref{fig2} the dashed line points to the starting state of the system and in the green boxes we define the different error states that are possible during the processing of a specific task.
\begin{enumerate}
	\item  The first two green boxes define only syntax errors of the first type that indicate a misunderstanding of the programming language and spelling mistakes in written commands. Within the correction graph, we can analyze several aspects that are part of the environment surrounding the error~\cite{r6}.
	\item  The number of mistakes made in implementing the program. A single execution of the code leads to the generation of $N$ in the number of errors, the set of $\{N\}$ of all errors committed leads to the current status~\cite{r8}.
	\item  The interval between the individual errors if there is a number of identical errors within a certain period. The interval can indicate whether the agent has experimented or obtained the correct syntactic construction or the state of the code over time has led to unrelated errors. In this way, it can be estimated to what extent the trained agent encounters difficulties with a single structure or a set of such.
	\item  The state of the code in the different stages of the erroneous assumption.
\end{enumerate}

\textbf{Conclusion}

One of the most challenging topics related to e-learning today is that of evaluating the work and competences of learners within the course. In this line of thought, the involvement of the teaching staff as well as of the supervisors and assistants responsible for conduct or successful facilitation must be taken into account. The answer to this problem will enable us to develop adequate tools that will give us more accurate information about the statistical status of each agent during the educational process as well as the overall commitment of course facilitators to the work they have done.

\textbf{Error classification as mechanism for sustainable educational process}

The sustainable development of our society is possible only through building mechanisms for an effective, optimized and predictive education process centered on the end-user of its education services as well as all tools needed for reaching this goal. The approach we suggest will give yet another instrument for the achievement of a modern and sustainable education. In order to support this process we offer a learning course around the problem at hand.

\textbf{Summer school course and requirements}

As part of a summer school a learning curriculum will be presented which will introduce to the relevant parties the aspects of the subject matter. In the structure of these sessions the following topics will be discussed:
\begin{itemize}
	\item  What is the user's behavior and how we can aid their digital registration, analysis and planning.
	\item  How to build an infrastructure and a medium which can monitor digitally registered actions.
	\item  Analysis and prediction for future activities.
	\item  Building neuron networks and teaching them with the goal to analyze repetitive activities.
\end{itemize}

With the goal of achieving the maximum possible number of discussed topics the students would need to have existing knowledge in areas having to do with object-oriented programming, artificial intelligence, including basic algorithms for searching and orientation as well as knowledge of ideas such as machine learning and neuron networks. It is preferred that the students have existing knowledge in programming languages such as Java / C\#  or C++.

\subsubsection{Acknowledgements} 

This paper acknowledges the support of the Erasmus+ Key Action 2 (Strategic partnership for higher education) project No. 2020-1-PT01-KA203-078646: ``SusTrainable -- Promoting Sustainability as a Fundamental Driver in Software Development Training and Education''.

The information and views set out in this paper are those of the author(s) and do not necessarily reflect the official opinion of the European Union. Neither the European Union institutions and bodies nor any person acting on their behalf may be held responsible for the use which may be made of the information contained therein.

\title{Adoption of Blockchain Technology for Tracking Students' Progress}
\author{Mihail Petrov\and Vladimir Valkanov}
\authorrunning{M. Petrov, V. Valkanov}
\institute{University Of Plovdiv ``Paisii Hilendarski'', \\ Faculty of Mathematics and Informatics, \\
24 Bulgaria Blvd., 4027 Plovdiv, Bulgaria \\
\email{mihailpetrov@uni-plovdiv.bg}\\
\email{vvalkanov@uni-plovdiv.bg}}
\maketitle
\begin{abstract}
With the increasing popularity of crypto currencies in contemporary economics the application of the blockchain technology in the control of processes other than payments is getting even more coverage. On an abstract level, the networks for transfer of financial assets can be equated to mechanisms for control of all information which solves problems having to do with information security and its completeness. In this current paper we will look at the use of blockchain for securing the information integrity in controlling educational processes.
\end{abstract}

\section{Introduction}

The occurrence of each event in the process of teaching through systems providing educational contents presupposes a change in the entire educational model. This educational model is directed both to a specific individual and to the entire educational process with the goal of following the learners' individual progress. 

For that reason it is important in every moment to track and store detailed characteristics in the form of a historical reference for the separate time nodes where a given event took place. An important moment in the examination of systems describing a given process' development is the element of trust in them. How can you be sure that the analysis of a given process was accurate if  you cannot trust the completeness and accuracy of the data while collecting it.

\section{What is the necessity for block chains in storing information}

Information security is an aspect encompassing all spheres of the modern industry, transportation of goods and services, providing medical help, education. In the digital communication era traditional educational hubs such as schools and universities are more frequently taking a backseat to alternative approaches in the form of electronic education. Information sources can now be found not just in traditional institutions but in professionals in the field of knowledge who are of interest to the learners.

The initiation and progress of the so-called MOOC (Massive Online Open Course)~\cite{rr1}, entirely realized with the use of modern communication  technologies, further blurs the borders between the classic university model and the classroom of the future.

Users in education are in the center of a vast ecosystem of professional courses offered by institutions of varying size and focus in classic universities and technological platforms. Education is even more accessible but the mechanisms for gatherin information for the students' activities during the learning process are limited or brought down to specific physical documents~\cite{rr2}. The end result of the education process is always a document showing the empirical measure of the student's efforts but hardly giving any insight into the steps towards that end result. 

In the context of traditional educational institutions, departments and faculties often use different information platforms to monitor the students' development within a particular course but this monitoring is usually isolated or ignores the student's previous development in their education. Additional knowledge is ignored and only the subject matter pertaining to the current knowledge and skills of the students is analyzed. Thus, extremely valuable information units are lost which does not allow for a detailed analysis of the learning process in order to answer the question whether the entire learning course leads to the desired results.

Two separate solutions can be offered for the solution of this problem:
\begin{itemize}
	\item Organizing the teaching curricula in a way which supposes direct progress based on course materials and results obtained in previous courses;
	\item	The use of a single platform which organizes and adapts the learning materials so that key components in the students' development are not omitted or depreciated -- which will be the focus of this current research. 
\end{itemize}

Let us look at an example of a standard distribution of programming disciplines in an academical or a training and learning environment. 
\begin{itemize}
	\item The students are studying at a minimum two programming languages in the form of introduction into programming;
	\item	The students are getting acquainted to the paradigm of object-oriented programming in the context of a given programming language;
	\item	The students are studying a conceptual set of approaches with mathematical nature having to do with solving a set of algorithm problems in the context of algorithms and data structures. 
\end{itemize}

Studying the given disciplines involves consecutive build-up of knowledges and skills in the following  areas:
\begin{itemize}
	\item Working with a programming language;
	\item	Working with concepts and paradigms;
	\item	Working with a logic-analytic algorithmic apparatus. 
\end{itemize}

The time period in which the active learning takes place can vary within the featured program but the minimum is four courses in two academic years. 

In that period the following inconsistencies are possible for the course material:
\begin{itemize}
	\item Lack of communication among groups of professors teaching the disciplines.
	\item	Unrealistic assessment of knowledges and skills acquired up to that moment.
\end{itemize}

The reasons for the aforementioned problems can be complex but it can also be difficult to make an informed choice for the current group of teachers if they do not have information about what happened during the learning process in previous disciplines and particularly if they lack historical information for reference.

One of the potential solutions is the active logging of activities from the students during their active interaction with the educational platform. 

\textbf{Blockchain} technology

The term blockchain is used in modern scientific technical literature to describe a broad set of concepts, including:
\begin{itemize}
	\item Describing a structure of data which is a set of cryptically interconnected information objects. The obects maintain constant relation and are not atomically separable, i.e.\ they cannot exist individually.
	\item	Technology for control of decentralized information clusters of data maintaining their consequence and allowing for precise tracking of the transaction nature of a given unit of historical data. 
\end{itemize}

The technological aspect of this paper is heavily influenced by the research activity of Satoshi Nakamoto in 2008 under the form of a platform for decentralized evaluation and transfer of digital monetary units known under its brand name Bitcoin~\cite{rr3}. The concept of the instrument is a direct juxtaposing of traditional architecture approaches for realization of centralized systems relying on expert server technologies for managing a cluster of peripheral processes. The nature of the given systems is usually private in character and serves trade interests of related companies with the goal of financial gains. Even though today this technology is mostly used with the goal of digital payments, its application is practically limitless and can be implemented in spheres of industry such as:
\begin{itemize}
	\item Control of deliveries and placement;
	\item	Monitoring vital signs and organization of medical files;
	\item	Managing letters of credit and other legal documents;
	\item	Managing and developing the education process.
\end{itemize}

Examples for successful uses of systems based on blockchain include:
\begin{itemize}
	\item Networks for transfer and trading virtual currencies such as Bitcoin~\cite{rr3}, Monero~\cite{rr15}, Eterium~\cite{rr16};
	\item	Decentralized systems for data storage such as Storj~\cite{rr13};
	\item	Systems for management of personal finances and monitoring investment instruments such as Neufund~\cite{rr14}.
\end{itemize}

\section{Advantages of blockchains to the traditional centralized models for data storage}

The main features which make the technology suitable for use for sensitive information can be characterized as follows:
\begin{itemize}
	\item A stable network eliminating the possibility for unwarranted modifications. The main concept of the blockchain network is that it is absolute. We can be certain that information added towards the blockchain will not get altered in the future or replaced by a third party.
	\item	Decentralization. The information is stored on a multitude of devices which makes it secure and practically protected from data damage as soon as there are active users within the system~\cite{rr5}.
\end{itemize}

Of course in its pure form the system has a number of weaknesses due to the theoretical model of its decentralized infrastructure. One of the main weaknesses comes from the algorithm for checking the block's completeness which, if not chosen properly, can lead to significant deterioration in the entire system's productivity~\cite{rr6}. Furthermore, if the correct model for compressing input data is not chosen, the cluster of information may become too sizeable to be adequately divided among all users in the network. 

\section{Integration of the blockchain in the context of an adaptive educational system}

The education is a sequential process, it starts and keeps going through one's life cycle. Formally, classic education can give you basic knowledges and skills which can then be developed through gathering professional experience and improving your skills by taking different courses, attending seminars or conferences as well as increasing your professional experience, practical experience and theoretical training.

To provide this learning, systems for electronic learning usually use systems for access to electronic education which provide a mechanism for interaction with the learning material, test solving, communication with the teachers and obtaining specific documents guaranteeing the end result of the education process~\cite{rr7}.

At the same time the teachers are still not aware of the students' additional activities in the online world which can give additional insight into the attempts to develop their potential in certain directions. For that reason it is necessary to possess a mechanism for complete monitoring of the students among seemingly disconnected platforms.

There are a  variety of internal student preparation and training systems operate at the University of Plovdiv. Examples of such systems are DELC1, DECL2, UniPlayground, their purpose is to support the learning process by utilizing the training tools needed and aggregating additional information providing a wide range of the possibilities of the teaching staff. For additional preparation, students most commonly use platforms such as Khan Academy to further improve their knowledge in Mathematics, Coursera for a variety of computer science courses, such as Artificial Intelligence, 

Introduction to Computer Science, Discrete Mathematics, Data Structures and Algorithms. The systems listed have self-contained modules to keep track of student development, such modules are:
\begin{itemize}
	\item Modules for solving test tasks, for intermediate and final evaluation;
	\item	Tracking system activity, reading articles, or checking homework;
	\item	Modules for reporting end activity, such as end value or end point asset;
	\item	Feedback modules from a teacher, colleague or student;
	\item	Modules for implementation and monitoring of team activities, group projects or supervision.
\end{itemize}

All these additional systems make it possible for students to improve their preparation, but the most important point is that they generate information, which quite often remains only within the system~\cite{rr8}.There is an indirect link between social networks and platforms, such as Coursera and Khan Academy, allowing the completion of a course to be tracked by providing a public link to access the platform and visualize information in the form of a certificate. This activity is carried out at the request of the participants in the program, and they have the opportunity to share all the information about their activity in the system. 

\textbf{UniChain private distributed blockchain platform}. In order to make the connection between the different systems, it is necessary to introduce an additional intermediate layer to aggregate and record information about the activities of students in the different platforms. This task is performed by an independent platform, UniChain, a decentralized cloud system, for managing the flow of resources between unconnected systems~\cite{rr9}. It is an anonymous, decentralized, system that stores a set of meta information regarding events occurring in educational systems. The events that have occurred are recorded, shared voluntarily by the participants in the courses and provide additional information to the consumers in the other courses.

UniChain is a decentralized sequence of logical blocks containing a timer and a collection of data that are linked together by a cryptographic identifier that uniquely indicates the direction between two linked blocks. The information that can be stored in each of the blocks is divided into the following categories:
\begin{itemize}
	\item Information on the intermediate result achieved;
	\item	Outcome information;
	\item	Meta information regarding an intermediate event;
\end{itemize}

Let's look at the three aspects in more detail. Intermediate result information is any result of an intermediate exam, or a final exam that accumulates any grade, the information about that result is processed by the system along with additional information about the activities that suggest that result.

At this stage, each block consists of the following identifiers
\begin{itemize}
  \item	The name of the discipline, the course;
	\item		Information on the subject on which the exam topic is administered;
	\item	Information about the exam material;
	\item		The end results.
\end{itemize}

The end result information includes the accumulated number of points or an empirical evaluation of the activities carried out so far, as well as verification of the certificate or diploma obtained from the accrediting platform, the student's results. Intermediate status meta information is an undefined and general area that makes it possible to record any important information resource, such as achieving consistent reading of a certain type of information, or attending classes regularly. This is a category of information that is specific to each individual platform and can support the student's overall information file depending on his or her activities.

\textbf{Architecture of an interface-based software system based on services}

Establishing active network communication, activity and wallet. To become part of the system, the user downloads a software portfolio provided by the educational institution. The portfolio plays the role of a mediator between the platforms the client visits and the UniChain network. At its core, the portfolio is a client built as a mobile or desktop application that has implemented interfaces to access the platforms that students visit~\cite{rr4}.

Users who are successfully logged into the UniChain system create a unique identifier for their activity and agree to share information from platforms that are integrated with the network.
\begin{figure}
\begin{center}
\includegraphics[width=0.6\textwidth]{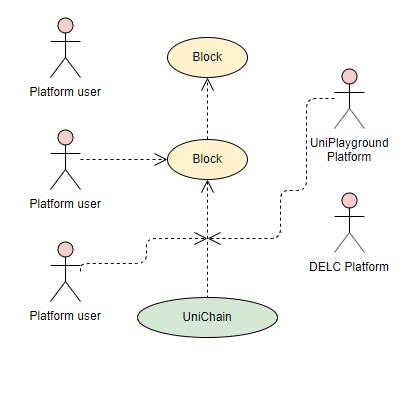}
\caption{Example plan for UniChain communication.}\label{2fig1}
\end{center}
\end{figure}

\textbf{Communication interfaces}

Communication between the various system components is based on interfaces. UniChain implements access to user information stored in the designated information resource flooding system and actively communicates when an event occurs. Event synchronization occurs either explicitly at the request of clients or with active communication between the system and the UniChain network.

UniChain is a blockchain application based on the Service oriented architecture model~\cite{rr11}. The main objective is to provide a set of services that can allow other connected systems to receive information from the network, with the express permission of the user. The active communication process enables integration with a new system to access information about all activities within existing systems. The interface allows users to categorize their activities without allowing them to modify the data.

\textbf{Storing of data}

A major problem with blockchain technologies is their scalability. Due to their decentralization, it is necessary to keep the files and information of the users consistent within the active devices. This often takes a huge amount of resources at a time when the network is gaining enough popularity. The storage of a 1MB file soon generates a huge amount of data that is stored within the active blocks. Such information may be a file, certificate, or other voluminous information.

UniChain solves the problem by keeping the files separate from their signature. The file storage signature is a cryptographically hashed meta data set indicating the file resource path stored on simple storage containers optimized for file resource storage. Only the identifiers that are being checked are matched to the active blocks on the chain blocks. The files containing the real resource are stored in a decentralized manner in an optimized cloud-based storage provider network. If an event occurs that invalidates the files, then their cryptographic signature will be compromised and no direct connection will be found between the replaced file and the recorded signature on the blockchain network.

\textbf{Discovering inconsistencies in the stored information (consensus algorithm)}

The blockchain loses its reliability when the data in it is commented on, so it is necessary to check to what extent the records in the individual blocks are positioned valid throughout the network while the distribution process is in progress. Conceptually the most popular verification algorithm is the Proof of work used on the Bitcoin network, but one of the major drawbacks is the concept of engaging resources competitively to be rewarded through some resource.

The UniChain Network does not provide a prize pool, so active resources will be verified for active parties using the Proof of Authority algorithm~\cite{rr10}. The concept behind this mechanism is that a small number of network members are authorized to verify the availability of an update, synchronizing blocks with all actors in the chain. Unlike the classic model, these participants will not be permanently involved in the process but will be selected at random depending on their activity. One of the private platforms selected at random will always be involved in the verification of the data.

\textbf{Conclusion}

In this article the architecture and main components of a private network for storing information for the students' achievements in the context of their electronic learning were presented. The technology for the application is based on the concept of storing information in the model of block chains warranting absoluteness and decentralization of the stored data with the goal of a high level of security and prevention of unwarranted activities. 

The project has the goal of providing up-to-date information on the results achieved in random educational institutions warranting transparency and an additional level of analysis in providing a quality information product to the learners~\cite{rr12}. The proposed model can be integrated in other areas of public electronic education while having in mind that changes having to do with such integration are dependent on the model for access to the system and the characteristics of the stored information.

\textbf{Blockchain and sustainability}

The tech society of today is becoming aware of one major problem which has to do with trust. Fake news, influences and information attacks are now an aspect of our everyday lives which eliminates our sustainable growth as a healthy thinking society. We believe that the toolset for the blockchains can get us closer to a world where trust is built through mutual consensus and leads only to tangible results based on achievements. With the topic being discussed we are trying to present one angle to this problem and propose steps towards its sustainable solution as well as a learning course to introduce the subject matter to the students.

\textbf{Summer school course and requirements}

As part of a summer school a learning curriculum will be presented which will introduce to the relevant parties the aspects of the subject matter. In the structure of these sessions the following topics will be discussed:
\begin{itemize}
	\item	What are blockchains and how their use helps us build trust among the users of information services.
	\item	How we can build a miniature model of a working blockchain and register in it all valid processes having to do with the transfer of information among the relevant parties.
	\item		Working with digital wallets and their integration in chains for service delivery.
	\item	Optimization of blockchains -- advantages and disadvantages of the modern algorithmic base on which the most recent 3 generations representing this industry were built.
\end{itemize}

With the goal of achieving the maximum possible number of discussed topics the students would need to have existing knowledge in object-oriented programming, preferably with languages such as Java / C\#  or C++. 

\subsubsection{Acknowledgements} 

This paper acknowledges the support of the Erasmus+ Key Action 2 (Strategic partnership for higher education) project No. 2020-1-PT01-KA203-078646: ``SusTrainable -- Promoting Sustainability as a Fundamental Driver in Software Development Training and Education''.

The information and views set out in this paper are those of the author(s) and do not necessarily reflect the official opinion of the European Union. Neither the European Union institutions and bodies nor any person acting on their behalf may be held responsible for the use which may be made of the information contained therein.

\end{document}